\documentclass[12pt]{article}
\usepackage{amsthm,amsmath,latexsym,amssymb,amsfonts,amscd}
\usepackage{graphics,lscape,fancyhdr,array,stmaryrd,euscript,wrapfig}
\usepackage{empheq,wrapfig}
\usepackage{verbatim,slashed}
\numberwithin{equation}{section}
\usepackage{hyperref,setspace}
\usepackage{tikz-cd}
\usepackage{mathrsfs}
\usepackage[numbers,sort&compress]{natbib}
\usepackage[nottoc]{tocbibind}

 \usepackage{xcolor}

\newcommand{\pl}{\partial}

\newcommand{\eps}{\varepsilon}

\newcommand{\fud}[2]{{}^{#1}{}_{#2}\,}
\newcommand{\fdu}[2]{{}_{#1}{}^{#2}\,}

\newcommand{\bry}{{{\bar{y}}}}

\newcommand{\brk}{{{\bar{k}}}}
\newcommand{\brq}{{{\bar{q}}}}
\newcommand{\brp}{{{\bar{p}}}}

\newcommand{\gad}{A'}

\newcommand{\ga}{A}

\newcommand{\besubeqs}{\begin{subequations}}
\newcommand{\esubeqs}{\end{subequations}}

\newcommand{\planeexp}{{e^{\pm i x^{CC'} k_C \brk_{C'}}}}
\newcommand{\ab}[1]{\langle{#1}\rangle} 
\newcommand{\rb}[1]{[{#1}]} 
\newcommand{\ub}[1]{\langle{#1}\rangle} 
\newcommand{\pb}[1]{[{#1}]} 

\newcommand{\propagator}[1]{\pi_{#1}}

\newcommand{\ybar}{\Bar{y}}

\newcommand{\Hbk}[2]{({#1} \ H \ {#2})}
\newcommand{\ebk}[2]{({#1} \ e \ {#2})}

\newcommand{\ampl}{\mathcal{A}}
\newcommand{\Lag}{\mathcal{L}}
\newcommand{\lbd}{\lambda}

\newcommand{\xcc}{\mathcal{X}}
\newcommand{\woo}{\mathcal{W}}
\newcommand{\voo}{\mathcal{V}}
\newcommand{\vooc}{\mathcal{V}_1}
\newcommand{\voco}{\mathcal{V}_2}
\newcommand{\vcoo}{\mathcal{V}_3}
\newcommand{\uoc}{\mathcal{U}}
\newcommand{\uocc}{\mathcal{U}_1}
\newcommand{\ucoc}{\mathcal{U}_2}
\newcommand{\ucco}{\mathcal{U}_3}

\newcommand{\slalg}[1]{\mathfrak{sl}(#1)}
\newcommand{\so}[1]{\mathfrak{so}(#1)}

\newcommand{\complex}{\mathbb{C}}

\newcommand{\pbar}{\Bar{p}}
\newcommand{\plbar}{\Bar{\partial}}
\newcommand{\kbar}{\Bar{k}}

\newcommand{\sigmabar}{\Bar{\sigma}}
\newcommand{\id}{1\!\!1}
\newcommand{\trace}{\mathrm{Tr}}
\begin{document}
\pagenumbering{gobble}
\hfill
\vskip 0.01\textheight
\begin{center}
{\Large\bfseries 
On amplitudes in Chiral Higher Spin Gravity}

\vspace{0.4cm}

\vskip 0.03\textheight
\renewcommand{\thefootnote}{\fnsymbol{footnote}}
Robin \textsc{Guarini}
\renewcommand{\thefootnote}{\arabic{footnote}}
\vskip 0.03\textheight

{\em Service de Physique de l'Univers, Champs et Gravitation, \\ Universit\'e de Mons, 20 place du Parc, 7000 Mons, 
Belgium}

\end{center}

\vskip 0.02\textheight

\begin{abstract}
We extract cubic interactions from the covariant equations of motion of Chiral Higher Spin Gravity and compute the corresponding amplitudes. These amplitudes are found to agree with earlier results obtained in the light-cone gauge. We also classify all possible cubic amplitudes that can be constructed from chiral fields arising in twistor theory. In addition, we derive propagators for arbitrary spin in the Feynman/Lorenz gauges. These propagators are then used to solve Berends--Giele recursion relations for a truncation of Chiral Higher Spin Gravity, which is a higher-spin extension of self-dual Yang--Mills theory. We confirm that all tree-level amplitudes vanish. 
\end{abstract}

\newpage
\tableofcontents
\newpage
\section{Introduction}
\label{sec:}
\pagenumbering{arabic}
\setcounter{page}{2}
Higher-spin gravities (HiSGRA) are toy models of quantum gravity that are expected to have improved UV-behaviour thanks to the presence of massless higher-spin fields, see e.g. \cite{Bekaert:2022poo}. However, having massless higher-spin fields in the spectrum is often in tension with the basic field-theory assumptions such as locality. As a result, there are only a handful of higher-spin gravities that ``behave'' as field theories in the sense of being local, but one has to sacrifice either propagating degrees of freedom or unitarity.   

For example, massless higher-spin fields do not have propagating degrees of freedom below four dimensions, which still leads to a plenty of interesting topological higher spin theories \cite{Blencowe:1988gj,Bergshoeff:1989ns,Campoleoni:2010zq,Henneaux:2010xg,Grigoriev:2020lzu,Pope:1989vj,Fradkin:1989xt,Grigoriev:2019xmp,Alkalaev:2014qpa,Alkalaev:2019xuv,Alkalaev:2020kut}. Some of these models can be extended with matter fields \cite{Sharapov:2024euk,Bekaert:2025azj}. Going up the dimension, in even dimensions there exist conformal higher-spin gravities \cite{Segal:2002gd,Tseytlin:2002gz,Bekaert:2010ky, Basile:2022nou}. The case of four dimensions is very special for massless fields, as the interactions of genuine higher-spin fields, $s>2$, reveal certain peculiarities that are not present in higher dimensions \cite{Bengtsson:1983pg,Bengtsson:1983pd,Bengtsson:1986kh,Metsaev:1991nb,Metsaev:1991mt,Bengtsson:2014qza, Conde:2016izb}. What is special is that a massless particle breaks down into two helicity states. 

A classical result, obtained first in the light-cone gauge \cite{Bengtsson:1983pg,Bengtsson:1983pd,Bengtsson:1986kh,Metsaev:1991nb,Metsaev:1991mt}, is that for any triplet of helicities $\lambda_{1,2,3}$ there is a unique cubic amplitude for $\lambda_1+\lambda_2+\lambda_3>0$ and $\lambda_1+\lambda_2+\lambda_3<0$. For example, in the former case the corresponding amplitude in the spinor-helicity language is 
\begin{align}\notag
   \mathcal{A}_{\lambda_1,\lambda_2,\lambda_3} \sim 
    [12]^{\lambda_1+\lambda_2-\lambda_3}[23]^{\lambda_2+\lambda_3-\lambda_1}[13]^{\lambda_1+\lambda_3-\lambda_2}\,,
\end{align}
The standard description of massless fields with spin is by a symmetric rank-$s$ tensor field $\Phi_{\mu_1\ldots \mu_s}$ \cite{Fronsdal:1978rb} fails to capture some of these interactions, including the most important gravitational and gauge interactions, see e.g. \cite{Bengtsson:2014qza, Conde:2016izb}. All $\lambda_1+\lambda_2+\lambda_3>0$ interactions are present in Chiral Higher Spin Gravity, which was originally found in the light-cone gauge \cite{Ponomarev:2016lrm} following the seminal works by Metsaev \cite{Metsaev:1991nb,Metsaev:1991mt}.  

In $d=4$ other opportunities to construct a covariant description of massless higher-spin fields can be explored to describe massless fields with spin. In principle, any spin-tensor $\Phi_{A(n),A'(m)}$ with $n+m=2s$ is good enough to carry the spin-$s$ degrees of freedom. The standard approach is to choose $m=n=s$, which is $\Phi_{\mu_1\ldots \mu_s}\sim \Phi_{A(s),A'(s)}$.  Alternatively \cite{Penrose:1965am}, one can take a $(2s,0)$-field $\Psi_{A(2s)}$ to describe negative helicity and $(0,2s)$-field $\Psi_{A'(2s)}$ to describe positive helicity, thereby treating them as independent. However, the range of interactions covered by local interactions of these fields is also very limited. The most promising approach, which originates from twistor theory \cite{Hughston:1979tq,Eastwood:1981jy,Woodhouse:1985id}, is to take $\Psi_{A(2s)}$ to describe one helicity and $\Phi_{A(2s-1),A'}$ to describe the opposite helicity. The latter field is a gauge field. 

Chiral Higher Spin Gravity was covariantized and extended to (anti)-de Sitter space in a series of works \cite{Sharapov:2022faa,Sharapov:2022wpz,Sharapov:2022awp,Sharapov:2022nps,Sharapov:2023erv}. Also its quantization has been studied and UV-finiteness was established at one-loop \cite{Ponomarev:2016lrm,Ponomarev:2017nrr,Skvortsov:2018jea,Skvortsov:2020wtf,Skvortsov:2020gpn,Tsulaia:2022csz}. While the equations involve infinitely many auxiliary fields, the dynamical fields that describe massless helicity fields are exactly $\Psi_{A(2s)}$ and $\Phi_{A(2s-1),A'}$. Some cubic amplitudes were already computed in \cite{Skvortsov:2022syz} as a consistency check and in this paper we complete them by extracting all cubic amplitudes from the equations of motion.

The main features of the free part of Chiral Higher-Spin Gravity are reviewed in the second section. The free equations of motion are discussed as well as their plane-wave solutions. The interactions are introduced in Section 3, which is mostly dedicated to the computation of cubic amplitudes by means of vertex operators. Section 4 is a discussion on the triviality of any potential quartic amplitude, based on the aforementioned vertex operators. The last section presents the application of the Berends--Giele recursion to a rather simple higher-spin theory: Higher-Spin Self-Dual Yang--Mills, which is a contraction of chiral higher-spin gravity. We show that all tree-level amplitudes vanish in this theory (except for the cubic ones). 

\section{Free fields}
\label{sec:}

The first linear (free) theory of massless higher-spin fields was established by Fronsdal \cite{Fronsdal:1978rb} by taking the massless limit of the Lagrangian worked out by Singh and Hagen \cite{Singh:1974qz}. In Fronsdal's theory, a massless spin-$s$ field is primarily described by a totally symmetric and double-traceless tensor $\Phi_{\mu_1...\mu_s}\equiv\Phi_{\mu(s)}$.\footnote{Denoting symmetric indices by the same letter is a common trick that will be extensively used throughout the text.} The Fronsdal equation of motion 
\begin{equation}
    \Box\Phi_{\mu(s)}-s\ \pl_\mu\pl\cdot\Phi_{\mu(s-1)}+\frac{s(s-1)}{2}\ \pl_\mu\pl_\mu\Phi'_{\mu(s-2)} = 0 \label{fronsdaleq}\ ,
\end{equation}
where $\pl\cdot\Phi$ and $\Phi'$ denote respectively the divergence and the trace of the field, is invariant under the gauge transformation 
\begin{equation}
    \delta\Phi_{\mu(s)}=\pl_\mu\xi_{\mu(s-1)} \ ,
\end{equation}
where the gauge parameter $\xi_{\mu(s-1)}$ is totally symmetric and traceless. 

Instead of working with $\Phi^{\mu_1...\mu_s}$, which lives in the \textit{symmetric representation}, our framework involves a gauge field, or gauge potential, $\Phi^{A(2s-1),A'}$ in the \textit{chiral representation}. The relevant equations and gauge transformations are
\begin{align}
    \partial\fud{A}{C'} \Phi^{A(2s-1),C'}&=0\,, && \delta \Phi^{A(2s-1),A'}=\partial^{AA'}\xi^{A(s-2)} \,.\label{firstphi}
\end{align}
The chiral field obeys a first-order differential equation, 
which, when contracted with $\pl_{AB'}$, yields the familiar second-order equation $\square \Phi=0$ by virtue of the identity $\pl_{AB'}\pl^{A}{}_{A'}=\tfrac{1}{2}\eps_{B'A'}\square$ (see Appendices \ref{appspin} and \ref{appdiff} for identities constraining spin-tensors). The opposite helicity field $\Psi^{A(2s)}$ obeys a first-order equation as well,
\begin{equation}
    \pl\fdu{B}{A'} \Psi^{BA(2s-1)}=0 \ . \label{firstpsi}
\end{equation}
These first-order equations of motion can be derived by varying the free Chiral Higher Spin Gravity action \cite{Krasnov:2021nsq}
\begin{align}\label{niceaction}
    S= \int \Psi^{A(2s)}\wedge H_{AA}\wedge \nabla \omega_{A(2s-2)}\,.
\end{align}
Here, $\Psi^{A(2s)}$ is interpreted as a zero-form and the one-form gauge field $\omega^{A(2s-2)}=\omega^{A(2s-2)}_\mu\ dx^\mu$ encodes $\Phi^{A(2s-1),A'}$, as we will see. $H^{AA}\equiv\tfrac{1}{2}H^{AA}_{\mu\nu}\ dx^{\mu}\wedge dx^\nu$ is a triplet of self-dual two-forms, which can be expressed in terms of the vierbein one-form $e^{AA'}\equiv e^{AA'}_\mu \ dx^\mu$ as $H^{AA}=e^{A}{}_{C'}\wedge e^{AC'}$  (see Appendices \ref{appspin} and \ref{appdiff} for more details); $\nabla= e^{AA'}\nabla_{AA'}$ is the Lorentz covariant derivative.\footnote{Acting on $\chi^{AA'}$, it writes $\nabla \chi^{AA'}=d\chi^{AA'}+\omega^A{}_B\ \chi^{BA'}+\omega^{A'}{}_{B'}\ \chi^{AB'}$.} It satisfies $\nabla^2\chi^A=0$. This property defines a class of manifolds called \textit{self-dual backgrounds} due to the fact that the self-dual half of the Weyl tensor $C^{ABCD}$ is forced to vanish. For our purposes, the chosen background will be simply Minkowski space, where $\nabla=d$. Nonetheless, it is worth mentioning that Chiral HiSGRA admits more general geometries than Minkowski or $(A)dS$, see e.g. \cite{Tran:2025yzd,Skvortsov:2025ohi,Skvortsov:2024rng} for some other exact solutions. This being said, let us relate this action to the previous equations. The one-form gauge field $\omega^{A(2s-2)}$, in which the gauge potential $\Phi^{A(2s-1),A'}$ is embedded, can be decomposed as follows.
\begin{align}
\omega^{A(2s-2)}\equiv e_{BB'}\Phi^{A(2s-2)B,B'}+e\fud{A}{B'}\Theta^{A(2s-3),B'}\,, 
\label{omPhiTheta}
\end{align}
where $\Phi$ and $\Theta$ are irreducible spin-tensors. The action enjoys the gauge transformation
\begin{equation}
    \delta \omega^{A(2s-2)}= d \xi^{A(2s-2)} +e\fud{A}{C'} \eta^{A(2s-3),C'}\,,
\end{equation}
where $\xi^{A(2s-2)}$ and $\eta^{A(2s-3),C'}$ are zero-forms. Of course, the differential part corresponds to the usual gauge transformation \eqref{firstphi}. The algebraic symmetry is present for $s>1$ and can be used to remove the $\Theta$-component in \eqref{omPhiTheta}, leaving us with the familiar $\Phi^{A(2s-1),A'}$. The action is gauge invariant under the algebraic symmetries thanks to the identity $H_{AA}\wedge e_{AB'}\equiv0$ (see Appendix \ref{appdiff}). In what follows we will impose the Lorenz gauge 
\begin{align}\label{Lorenzchir}
    \pl_{BB'}\Phi^{A(2s-2)B,B'}&=0
\end{align}
to make the calculations easier. Despite not completely fixing the gauge, this condition removes $2s-1$ components from $\Phi$, thereby reducing the number of components from $4s$ to $2s+1$. This matches the number of components carried by the field $\Psi$, as the number of components of a rank-$n$ totally symmetric tensor is $n+1$ in dimension 2. Therefore, the kinetic term is invertible. 

The main message of this introduction to free Chiral HiSGRA is that, instead of one single gauge field obeying a second-order equation and containing both helicities, the theory has two independent fields obeying first-order equations. Each field encodes one particular helicity. The following paragraphs presents a standard way of describing the free equations.

\subsection{Free Chiral HiSGRA as FDA system}

The free equations \eqref{firstphi} and \eqref{firstpsi} can be written in an FDA form (Free Differential Algebra).\footnote{FDA was introduced by Sullivan \cite{Sullivan77} and later adopted to study SUGRA \cite{vanNieuwenhuizen:1982zf,DAuria:1980cmy} and to study the problem of interactions of higher-spin fields \cite{Vasiliev:1988sa}. } This requires the addition of (infinitely-many) auxiliary fields. Starting from the variational equations
\begin{align}\label{variateq}
    &\nabla\Psi^{A(2s)}\wedge H_{AA}=0\,, && &H_{AA}\wedge\nabla\omega_{A(2s-2)}=0 \ ,
\end{align}
the derivatives of the physical fields can be parametrized by certain spin-tensors in such a way that the previous equations are automatically satisfied. For example, the first equation is satisfied if $\nabla\Psi^{A(2s)}=e_{BB'}\Psi^{A(2s)B,B'}$. The $\nabla$ of these new fields can be parametrized by other fields, etc... Finally, the result is a system of equations with a recursive pattern \cite{Skvortsov:2022syz} 
\begin{subequations} \label{FDAnabla}
    \begin{align}
    &\nabla \omega^{A(n-i),A'(i)} = e^A{}_{B'}\wedge\omega^{A(n-i-1),A'(i)B'}\, ,\label{FDAnablaOmOm}\\
    &\nabla \omega^{A'(n)} =  H_{B'B'} C^{A'(n)B'B'}\, ,\label{FDAnablaOmC}\\
    &\nabla C^{A(k),A'(n+k+2)} =  e_{BB'} C^{BA(k),B'A'(n+k+2)}\, ,\label{FDAnablaC}\\
    &\nabla \Psi^{A(n+k+2),A'(k)} = e_{BB'} \Psi^{BA(n+k+2),B'A'(k)} \, . \label{FDAnablaPsi}
    \end{align}
\end{subequations}
In general, equations of motion indicate what vanishes. Instead, the philosophy of FDA aims to answer the question “What does not vanish ?". Indeed, the derivative of the field on the left-hand side is parametrized by a new field on the right-hand side of each equation. In principle, there is no reason to do that for non higher-spin theories since the infinitely many auxiliary fields introduced to parameterized higher derivatives of the dynamical fields would be of little use in, say, Yang--Mills theory and gravity.\footnote{Finding the FDA formulation for ordinary theories is a challenge, see e.g. \cite{Skvortsov:2022unu,Misuna:2024dlx,Misuna:2024ccj}. } However, in HiSGRA higher-spin interactions contain higher derivatives and what the FDA-equations will do at the nonlinear level is to parameterize higher-derivatives in a higher-spin covariant way and to immediately use them to construct interaction vertices. Another aspect where FDAs can have an advantage is when an infinite-dimensional symmetry is present, which is our case as well.

It is most convenient to replace the numerous fields with free indices by \textit{generating functions}. For instance, $\Psi^{A(n+k+2),A'(k)}$ and $C^{A(k),A'(n+k+2)}$ can both be packed into a single generating function $C(y,\bry)$. 
\begin{equation}
    C(y,\ybar)=\sum_{n,m}\tfrac{1}{n!m!}\ C_{A(n),A'(m)}\ y^A\ldots y^A\ybar^{A'}\ldots \ybar^{A'} \, .
\end{equation}
In particular, this function contains the fields $C^{A(k),A'(k)}$ required to describe a scalar field, which is necessarily present in Chiral Theory. All the one-forms can be stored in 
\begin{equation}
    \omega(y,\ybar)=\sum_{n,m}\tfrac{1}{n!m!}\ \omega_{A(n),A'(m)}\ y^A\ldots y^A\ybar^{A'}\ldots \ybar^{A'} \, .
\end{equation}
A small remark must be made at this point. In this paper, all fields are bosonic. This implies $n+m$ is always even. When working with generating functions, contraction and symmetrization of indices are implemented by operations on ``twistor'' variables $y^A$ and $\ybar^{A'}$. Consider two generating functions $\alpha(y_1,\ybar_1)$ and $\beta(y_2,\ybar_2)$ which have degree 1 in each twistor variable. Symmetrizing unprimed indices and contracting primed indices can be realized by means of simple differential operators. Indeed, for example, 
\begin{equation*}
    (y^A\pl_{1A})(y^B\pl_{2B})(\plbar_1^{A'}\plbar_{2A'})\ \alpha(y_1,\ybar_1) \beta(y_2,\ybar_2)=\alpha_{A,}{}^{A'}\beta_{A,A'} \, .
\end{equation*}
 Generating functions allow for a more compact way of writing equations \eqref{FDAnablaOmOm}, \eqref{FDAnablaOmC}, \eqref{FDAnablaC} and \eqref{FDAnablaPsi}, which now read \cite{Skvortsov:2022syz}\footnote{These equations are very close to \cite{Vasiliev:1986td}. Indeed, the spectrum of FDA fields depends on the degrees of freedom one wishes to describe, rather than on what kind of Lorentz-covariant dynamical field is chosen. In particular, $\omega^{A(s-1),A'(s-1)}$ contains an equivalent of the Fronsdal field, despite the fact that it is the chiral fields $\Phi^{A(2s-1),A'}$ and $\Psi^{A(2s)}$ that are dynamical. }
\begin{align}\label{linearizeddata}
    \nabla\omega &= e^{BB'}y_{B} \pl_{B'} \omega +H^{B'B'} \pl_{B'}\pl_{B'}C(y=0,\bry)\,,& 
    \nabla C&= e^{BB'}\pl_B \pl_{B'} C\,.
\end{align}

\subsection{Plane waves}
\label{sec:}
In order to compute amplitudes we need to know plane-wave solutions. We denote the physical momenta by $k^\mu$. As we consider only massless fields, all momenta satisfy $k^2=0$, and, hence, $k^{AA'}=k^A \brk^{A'}$, where $k^{AA'}=\sigma^{AA'}_\mu k^\mu$ (see Appendix \ref{appspin}). A convenient gadget to parameterize the gauge ambiguity of solutions is to choose a reference spinor $q^A$, $\brq^{A'}$. Just for completeness, let us start with the plane-wave solutions within the symmetric approach.
\begin{align}
\begin{aligned}
\Phi_{\ga(s),\gad(s)}(k)&=
    \frac{k_{\ga_1}...k_{\ga_s}\brq_{\gad_1}...\brq_{\gad_s}}{( \brq\brk)^s} \planeexp \Phi_{-s}(k)+\\
    &+\frac{q_{\ga_1}...q_{\ga_s}\brk_{\gad_1}...\brk_{\gad_s}}{(qk)^s} \planeexp \Phi_{+s}(k)\,,
\end{aligned}
\end{align}
where $(qk)\equiv q^Ak_A$ and $(qk)\equiv \brq^A\brk_{A'}$. The solution contains both helicities as it should. However, taking the first curl and projecting it on an irreducible component eliminates one of the helicities, e.g.\footnote{This does not happen in the presence of the cosmological constant, see \cite{Skvortsov:2022wzo}. }
\begin{align}\label{helicityout}
    \pl\fdu{\ga}{B'} \Phi_{\ga(s),\gad(s-1)B'}&\sim \frac{k_{\ga_1}...k_{\ga_{s+1}}\brq_{\gad_1}...\brq_{\gad_{s-1}}}{(\brq\brk)^{s-1}} \planeexp \Phi_{-s}(k) \,.
\end{align}
Taking more curls leads to a family of fields
\begin{align}\notag
    \Phi_{\ga(s+n),\gad(s-n)} &= \pl\fdu{\ga}{B'}...\pl\fdu{\ga}{B'} \Phi_{\ga(s),\gad(s-n) B'(n)}\sim\frac{k_{\ga_1}...k_{\ga_{s+n}}\brq_{\gad_1}...\brq_{\gad_{s-n}}}{(\brk \brq)^{s-n}} \planeexp \Phi_{-s}(k)\,.
\end{align}
that ends with the gauge invariant field-strength
\begin{align}\label{allmycurls}
    \Psi_{\ga(2s)} &\equiv \pl\fdu{\ga}{B'}...\pl\fdu{\ga}{B'} \Phi_{\ga(s), B'(s)}\sim k_{A_1}...k_{A_{2s}} \planeexp \Phi_{-s}(k)\,.
\end{align}
This is exactly the spin-tensor field introduced previously if we take it to be an independent dynamical field instead of relating it to the gauge potential. Similarly, one can introduce $\Psi_{A'(2s)}$ and the two helicities turn out to be completely disentangled. It is also obvious that the only on-shell nontrivial derivatives of these two fields are
\begin{align}\label{higherder}
    \Psi_{\ga(2s+k),\gad(k)}&= \pl_{\ga\gad} ...\pl_{\ga\gad}\Psi_{\ga(2s)}\,, & \Psi_{\ga(k),\gad(2s+k)}&= \pl_{\ga\gad} ...\pl_{\ga\gad}\Psi_{\gad(2s)}\,,
\end{align}
which explains the spectrum of $C(y,\bry)$ that was introduced before. Yet another dynamical field we need is gauge potential $\Phi^{A(2s-1),A'}$, whose plane wave representation reads
\begin{align}
    \Phi^{A(2s-1),A'}&= \frac{1}{(qk)^{2s-1}} q^{A} \ldots q^{A} \brk^{A'}\planeexp \Phi_{+s}(k)
    \label{Phiindices}
\end{align}
Note, however, that it is not the last but one member of the $\Phi_{\ga(s-n),\gad(s+n)}$-family, just one derivative away from the gauge invariant field-strength $\Psi^{A'(2s)}$. It is right the opposite, given the number of the reference spinors it contains. One can construct a family of ``less and less gauge non-invariant'' curls
\begin{align}
   \Phi^{A(2s-1-n),A'(n+1)}&= \pl\fdu{B}{A'}\ldots \pl\fdu{B}{A'} \Phi^{A(2s-1-n)B(n),A'}\\
   &= \frac{1}{(qk)^{2s-1-n}} q^{A(2s-1-n)} \brk^{A'(n+1)}\planeexp \Phi_{+s}(k)
\end{align}
which are components of the master field $\omega$. The family ends up with the gauge invariant field strength $C^{A'(2s)}$, after which the on-shell nontrivial derivatives are constructed the same way as for $\Psi^{A'(2s)}$ above.

Within the twistor approach, it is the pair $\Phi^{A(2s-1),A'}$ and $\Psi^{\ga(2s)}$ that are taken to be independent dynamical fields. The one-form gauge potential for the former one is
\begin{align}
    \omega^{A(2s-2)}&=  \frac{1}{(qk)^{2s-1}} \ebk{q}{\kbar} \,q^{A(2s-1)} \ e^{\pm kx\kbar}\ \Phi_{+s}(k)\ ,
\end{align}
where $\ebk{q}{\kbar}$ and $q^{A(2s-1)}$ are shorthand notations, respectively, for $e^{BB'}q_{B}\brk_{B'}$ and $q^A\ldots q^A$ ($2s-1$ times). From now on, we write the exponential as $\exp(\pm kx\kbar)$, $x^{CC'}$ being anti-Hermitian. With such notations, the plane-wave solution for the other field is 
\begin{equation}
    \Psi^{A(2s)}=\tfrac{1}{2}\ k^{A(2s)}\ e^{\pm kx\kbar} \ \Phi_{-s}(k) \, .
\end{equation}
Now, we can pack the dynamical fields together with the on-shell nontrivial derivatives thereof into two generating functions and drop the external labels $\Phi_{\pm s}(k)$ as well as $\planeexp$, we get
\begin{align}
    \omega(x|y,\bry)&=e^{BB'}\frac{q_{B}\brk_{B'}}{qk+yq}e^{\bry \brk}\ e^{\pm kx\kbar}\ , & C(x|y,\bry)=\tfrac{1}{2}\exp(yk+\bry\bar{k})\ e^{\pm kx\kbar} \ .
    \label{solgenfn}
\end{align}
These free solutions contain fields\footnote{Here, the name “fields" refers to both physical fields and their derivatives which are parametrized by spin-tensors having the same index structure.} of all (integer) spins.
An important ingredient of the free theory must be introduced before discussing the interactions. These are the polarization (spin-)tensors, which can be deduced from the equations \eqref{firstphi} and \eqref{firstpsi}, written in momentum space (see \cite{Krasnov:2021nsq}). The positive and negative polarisation tensors are respectively
\begin{equation}
    \eps^+_{A(2s-1),A'}(k) = M^{s-1}\ \frac{q_{A_1} \cdots q_{A_{2s-1}} k_{A'}}{(qk)^{2s-1}} \, ,
    \label{pospol}
\end{equation}
and
\begin{equation}
    \eps_{A(2s)}^{-}(k) = M^{1-s}\ k_{A_1} \ldots k_{A_{2s}}\, .
    \label{negpol}
\end{equation}
The dimensionful ``mass'' parameter $M$ will not be explicitly considered in the calculations. Its purpose is to ensure that positive and negative-helicity have respective mass dimensions 0 and 1, for any spin.

\section{Amplitudes}
\label{sectionAmpl}

In principle, one can classify all possible vertices in terms of the dynamical fields $\Phi^{A(2s-1),A'}$ and $\Psi^{\ga(2s)}$ themselves, or in terms of the on-shell nontrivial derivatives thereof. To simplify a bit the problem, it can be shown that all on-shell vertices can be written in terms of the master fields $\omega$ and $C$, with only $\wedge$-products among $\omega$, see e.g. \cite{Grigoriev:2020lzu} for the proof in $3d$ that can readily be extended to any dimension. Conceptually, since the master fields $\omega$ and $C$ contain the complete set of dynamical fields together with their on-shell nontrivial derivative, any local interaction of the dynamical fields is a no-derivative interaction of the master fields. Any $n$-point contact vertex and, hence, the corresponding amplitude can be represented as a poly-differential operator acting on a number of plane waves. More explicitly, an amplitude or, better say, a Lagrangian vertex four-form integrand can take the following expressions. 
\begin{align}
    \mathcal{A}_n&= 
    \left\{\begin{aligned}
        &\mathcal{A}(p_i,\brp_i)\, H_{AA}\wedge H^{AA}\, C(y_1,\bry_1)\ldots C(y_n,\bry_n)\\
        &\mathcal{A}_{AA'}(p_i,\brp_i)\, \hat{h}^{AA'}\wedge  \omega(y_1,\bry_1)\,C(y_2,\bry_2)\ldots C(y_n,\bry_n)\\
        &\mathcal{A}_{AA}(p_i,\brp_i)\,H^{AA} \wedge \omega(y_1,\bry_1)\wedge \omega(y_2,\bry_2)\,C(y_3,\bry_3)\ldots C(y_n,\bry_n)+c.c.\\
        &\mathcal{A}_{AA'}(p_i,\brp_i)\, e^{AA'}\wedge \omega(y_1,\bry_1)\wedge \omega(y_2,\bry_2)\wedge \omega(y_3,\bry_3)\,C(y_4,\bry_4)\ldots C(y_n,\bry_n)\\
        &\mathcal{A}(p_i,\brp_i)\, \omega(y_1,\bry_1)\wedge \omega(y_2,\bry_2)\wedge \omega(y_3,\bry_3)\wedge \omega(y_4,\bry_4)\,C(y_5,\bry_5)\ldots C(y_n,\bry_n)
    \end{aligned}\right|_{y_i=\bry_i=0}
    \label{listpossiblesampl}
\end{align}
where $p_i^A=\pl_{y_i}^{A}$ and $\brp_i^{A'}=\pl_{\bry_i}^{A'}$.  
Here we listed all possible combinations of $\omega$ and $C$ that build up a four-form that can be considered as an $n$-point (on-shell) interaction. There are three main requirements to be imposed on the poly-differential operators $\mathcal{A}(p_i,\brp_i)$: a) they have to be Lorentz invariant, hence, depend on the scalar products of the spinors; b) they have to admit Taylor expansion in $p_i$ and $\brp_i$ since the vertices have to represent certain contractions of indices of the master fields; c) they must be gauge invariant, which means that the vertices must not depend on the reference spinor $q$.  In the next part of the text, we explore all possible cases one by one.

\subsection{Three-point amplitudes}
\label{sec:}
It is well-known \cite{Bengtsson:1986kh,Benincasa:2011pg} that for any given three helicities $\lambda_{1,2,3}$, $\Lambda=\sum_i\lambda_i$, there is a unique nontrivial vertex and, hence, the amplitude for $\Lambda\neq0$: 
\begin{align}\label{genericV}
   \mathcal{A}_{\lambda_1,\lambda_2,\lambda_3}=V_{\lambda_1,\lambda_2,\lambda_3}\Big|_{\text{on-shell}} \sim \left\{\begin{aligned}
         &[12]^{\lambda_1+\lambda_2-\lambda_3}[23]^{\lambda_2+\lambda_3-\lambda_1}[13]^{\lambda_1+\lambda_3-\lambda_2}\,, && \Lambda>0\\
        &\langle 12\rangle^{\lambda_1+\lambda_2-\lambda_3}\langle23\rangle^{\lambda_2+\lambda_3-\lambda_1}\langle13\rangle^{\lambda_1+\lambda_3-\lambda_2}\,,&& \Lambda<0\,.
   \end{aligned}\right.
\end{align}
A more standard, spinor-helicity, notation has been introduced. The unprimed and primed momentum spinor products are denoted respectively by angle and square brackets.\footnote{An exception to this rule will be made for products involving twistor variables, such as $yq\equiv y^Aq_A$.}

What defines any theory at the cubic level is the spectrum of fields and the relative strength of the cubic interactions. The Chiral HiSGRA contains fields of all spins and the dependence of the coupling constants is fixed by Lorentz invariance to be \cite{Metsaev:1991mt,Metsaev:1991nb,Ponomarev:2016lrm}:
\begin{align}\label{eq:magicalcoupling}
    V_{\text{Chiral}}&= \sum_{\lambda_1,\lambda_2,\lambda_3}  C_{\lambda_1,\lambda_2,\lambda_3}V_{\lambda_1,\lambda_2,\lambda_3}\,, && C_{\lambda_1,\lambda_2,\lambda_3}=\frac{\kappa\,(l_p)^{\lambda_1+\lambda_2+\lambda_3-1}}{\Gamma(\lambda_1+\lambda_2+\lambda_3)}\,.
\end{align}
Here $l_p$ is the Planck length. This parameter is justified by dimensional purposes. On the one hand, $V_{\lambda_1,\lambda_2,\lambda_3}$ has mass dimension $\Lambda$ and $(l_p)^{\lambda_1+\lambda_2+\lambda_3-1}$ has mass dimension $1-\Lambda$. On the other hand, n-point amplitudes must have mass dimension $d-n$, $d$ and $n$ being respectively the dimension of space-time and the number of external legs.

As mentioned previously, scattering amplitudes must not depend on the reference spinor $q$ that appears in the expression of the free solution for the gauge field. This spinor will be removed by using momentum conservation, which for a cubic amplitude writes as 
\begin{equation}
    k^A_1 \brk^{A'}_1+k^A_2 \brk^{A'}_2+k^A_3 \brk^{A'}_3=0 \, .
\end{equation}
In practice, the index-free identity obtained by sandwiching the conservation law with $q_A$ on the left and one of the $\brk^i_{A'}$ on the right is more handy. Using shorthand notation $\ub{k_ik_j}\rightarrow\ub{ij}$ (and $\pb{\brk_i\brk_j}\rightarrow\pb{ij}$), we have the following momentum conservation constraints.
\begin{equation}
\begin{aligned}
    \ub{q2}[21]+\ub{q3}[31]&=0\, ,\\  
    \ub{q1}[12]+\ub{q3}[32]&=0\, ,\\ 
    \ub{q1}[13]+\ub{q2}[23]&=0 \, .
\end{aligned}
\label{momconsqk}
\end{equation}
By squaring the conservation law $k_1+k_2=-k_3$ and the similar ones we get $\ub{ij}\pb{ij}=0$, which just means that there is no three-point scattering of spinning particles for the real kinematics (since $i\rangle$ and $i]$ are complex conjugate of each other). However, the amplitudes do not vanish in the split signature $(2,2)$ or for complex kinematics. Therefore, it is assumed that $\brk$ is not necessarily the complex conjugate of $k$.\footnote{This is tantamount to considering complex four-vector components $k^\mu$.} $\ub{ij}\pb{ij}=0$ also means that the amplitudes cannot be proportional to  $\ub{ij}\pb{ij}$. As a result, they have to be made either from $\ub{ij}$ or from $\pb{ij}$, i.e. to be (anti)-holomorphic. The following paragraphs present the analysis of amplitudes that may be constructed in our higher higher-spin theory, following the list of interactions presented in \eqref{listpossiblesampl}. The overall sign or numerical factor will not be taken into account. Only the general structure matters. A generic cubic amplitude computed in position space is, schematically, given by
\begin{equation*}
    \ampl_3\propto\int \mathcal{A}(p_i,\brp_i)\times\Omega\times \varphi \, ,
\end{equation*}
where $\varphi$ represents the three fields and $\Omega$ is a differential form such that the total degree of the integrand is four. After extracting the vierbein form from each gauge field and using the identities from Appendix \ref{appdiff}, the expression becomes
\begin{equation*}
    \ampl_3\propto\int H\wedge H \ \mathcal{A}(p_i,\brp_i)\times \Tilde{\varphi} \ e^{\pm k_{1}x\kbar_{1}\pm k_{2}x\kbar_{2}\pm k_{3}x\kbar_{3}} \, .
\end{equation*}
$H\wedge H$ is the volume form, and $\Tilde{\varphi}$ is a zero-form which contains $\Phi$ instead of $\omega$. Integrating-out the space-time dependence leads to a delta distribution $\delta^4(k_1+k_2+k_3)$ which expresses momentum conservation. Finally, the amplitude looks like
\begin{equation}
    \ampl_3\propto \mathcal{A}(p_i,\brp_i)\times \Tilde{\varphi}  \ \delta^4(k_1+k_2+k_3) \, ,
\end{equation}
where $\mathcal{A}(p_i,\brp_i)$ is some operator contracting the indices of the free fields contained in $\Tilde{\varphi}$. Bearing in mind that momentum conservation holds, the delta function can be dropped for notational convenience. In what follows, the free solutions \eqref{solgenfn} will be used to generate all the possible contractions.  First, $\varphi$ will be a zero-form. Then, the other cases will result from the repetitive increase of the degree of $\varphi$ by 1.  

\paragraph{$\boldsymbol{CCC}$.} This is the simplest class of vertices since there is no dependence on the reference spinor and, hence, any vertex is automatically gauge-invariant. Given that the three-point vertices must be (anti)-holomorphic, poly-differential operator $\mathcal{A}(p_i,\brp_i)$ has to depend either on $p_i$ or $\brp_i$, but not both. Therefore, we get only $+++$ or $---$ amplitudes: 
\begin{align}
    &\mathcal{A}_{+s_1,+s_2,+s_3} &&\mathcal{A}_{-s_1,-s_2,-s_3}
\end{align}
These amplitudes correspond to abelian interactions. Consider a triplet of fields $C^{A(\alpha),A'(\beta)}$, $C^{A(\gamma),A'(\delta)}$ and $C^{A(\epsilon),A'(\zeta)}$. where $\beta-\alpha=2\lambda_1$, $\delta-\gamma=2\lambda_2$ and $\zeta-\epsilon=2\lambda_3$. A priori, each field could carry either of the helicities. 
The amplitude has the following structure,
\begin{equation}
    \mathcal{A}_3\propto \ab{12}^{a} \ \ab{23}^{b} \ \ab{31}^{c}\ \rb{12}^{d} \ \rb{23}^{b} \ \rb{31}^{c} \, ,
\end{equation}
where all the exponents are positive. A quick comparison between the fields and the last expression allows to relate $\alpha$, $\beta$, $\gamma$, $\delta$, $\epsilon$, $\zeta$ and $a$, $b$, $c$. These parameters obey the following constraints:
\begin{equation}
    \begin{cases}
        a=\frac{\alpha+\gamma-\epsilon}{2}\, , \\
        b=\frac{-\alpha+\gamma+\epsilon}{2}\, , \\
        c=\frac{\alpha-\gamma+\epsilon}{2}\, ,
    \end{cases}\quad ; \qquad
\begin{cases}
        d=\frac{\beta+\delta-\zeta}{2}\, , \\
        e=\frac{-\beta+\delta+\zeta}{2}\, , \\
        f=\frac{\beta-\delta+\zeta}{2}\, .
    \end{cases}
\end{equation}
As mentioned previously, two cases have to be treated, following the on-shell condition. The first one is $\ab{12}=\ab{23}=\ab{31}=0$, which implies $a=b=c=0$. As a consequence, $\alpha=\gamma=\epsilon=0$ and $\beta=2\lambda_1$, $\delta=2\lambda_2$ and $\zeta=2\lambda_3$. For each field, the number of primed indices is directly related to its helicity, which is positive. Therefore, the amplitude takes the following form,
\begin{equation}
    \mathcal{A}_3\propto \rb{12}^{\lambda_1+\lambda_2-\lambda_3} \ \rb{23}^{-\lambda_1+\lambda_2+\lambda_3} \ \rb{31}^{\lambda_1-\lambda_2+\lambda_3} \, .
\end{equation}
In the other case, the amplitude only involves angle brackets and $\alpha=-2\lambda_1$, $\gamma=-2\lambda_2$ and $\epsilon=-2\lambda_3$. Here, the helicities are all negative ($\lbd_i=-s_i$). The result is 
\begin{equation}
    \mathcal{A}_3\propto \ab{12}^{-\lambda_1-\lambda_2+\lambda_3} \ \ab{23}^{\lambda_1-\lambda_2-\lambda_3} \ \ab{31}^{-\lambda_1+\lambda_2-\lambda_3} \, .
\end{equation}
In both situations, the spins obey some triangular inequalities:
\begin{equation}
    \begin{cases}
        s_1+s_2 \geqslant s_3 \, ,\\
        s_1+s_3 \geqslant s_2 \, ,\\
        s_2+s_3 \geqslant s_1 \, .
    \end{cases}
\end{equation}

\paragraph{$\boldsymbol{\omega CC}$.}

Consider now a triplet of fields $\omega^{A(2s_1-2-k),A'(k)}$, $C^{A(\alpha),A'(\beta)}$, and $C^{A(\gamma),A'(\delta)}$. Of course, $\alpha$, $\beta$, $\gamma$ and $\delta$ are subject to the constraints $\beta-\alpha=2\lambda_2$, $\delta-\gamma=2\lambda_3$. The amplitude must have the following structure,
\begin{equation}
    \mathcal{A}_3\propto \frac{\ab{q2}^{a}  \ab{q3}^b}{\ab{q1}^{2s_1-1-k}} \ab{23}^{c}\ \rb{12}\  \rb{12}^{d} \ \rb{23}^{e} \ \rb{31}^{f} \ ,
\end{equation}
where all the exponents are positive. It is worth stressing that the gauge field contains a one-form vierbein contracted with a reference momentum $q_A$ and a primed spinor $\kbar_{1A'}$. Therefore, the total number of reference momenta carried by $\omega$ is $2s_1-1-k$. Moreover, the single momentum $\kbar_{1A'}$ has been contracted with $\kbar_{2A'}$.\footnote{It could be contracted with $\kbar_{3A'}$ as well, keeping the final result unaffected.} Hence the presence of the extra bracket '$[12]$'. The next step is to express the exponents in terms of $s_1$, $k$, $\alpha$, $\beta$, $\gamma$ and $\delta$. 
\begin{equation}
    \begin{cases}
        a=\frac{2s_1-1-k+\alpha-\gamma}{2}\, , \\
        b=\frac{2s_1-1-k-\alpha+\gamma}{2}\, , \\
        c=\frac{-2s_1+1+k+\alpha+\gamma}{2}\, ,
    \end{cases}\quad ; \qquad
\begin{cases}
        d=\frac{k+\beta-1-\delta}{2}\, , \\
        e=\frac{-k+\beta-1+\delta}{2}\, , \\
        f=\frac{k-\beta+1+\delta}{2}\, .
    \end{cases}
    \label{systeqomcc}
\end{equation}
The amplitude being a gauge-invariant quantity, it cannot depend on $q$. The reference spinor can be cancelled from the expression by using momentum conservation \eqref{momconsqk}. For example, 
\begin{equation*}
    \frac{\ab{q2}}{\ab{q1}}=\frac{[31]}{[23]}  \, .
\end{equation*}
Then, the amplitude becomes
\begin{equation}
    \mathcal{A}_3\propto \rb{12}^{s_1+\lambda_2-\lambda_3} \ \rb{31}^{s_1-\lambda_2+\lambda_3} \ \rb{23}^{-s_1+\lambda_2+\lambda_3+(\alpha+\gamma+k+1-2s_1)/2}\ \ab{23}^{(\alpha+\gamma+k+1-2s_1)/2} \, ,
\end{equation}
where the various conditions constraining the powers of the spinor brackets have been used to make the helicities appear. The final answer is obtained by using the special three-point kinematics. If angle brackets vanish, i.e. if $\alpha+\gamma+k+1-2s_1=0$, the result is
\begin{equation}
    \mathcal{A}_3\propto \rb{12}^{s_1+\lambda_2-\lambda_3} \ \rb{23}^{-s_1+\lambda_2+\lambda_3}\ \rb{31}^{s_1-\lambda_2+\lambda_3}\ .
\end{equation}
However, if one imposes the other condition, there is an inconsistency with the initial data. Indeed, a consequence of $\rb{12}=\rb{23}=\rb{31}=0$ is $s_1=0$, which is not valid since $\omega(y,\ybar)$ contains fields of all integer spins greater or equal to 1. The presence of the gauge field $\omega$ from Chiral Theory (with only positive helicity, $\lambda_1=s_1$) enforces the sum of helicities to be greater than 0. In fact, by injecting the constraint $\alpha+\gamma+k+1-2s_1=0$ in \eqref{systeqomcc}, the helicities satisfy 
\begin{equation}
    \begin{cases}
        s_1+\lambda_2+\lambda_3 \geqslant \text{max}\{k+1,\beta,\delta+1\}\, , \\
        s_1-\lambda_2+\lambda_3 \geqslant \alpha\, ,\\
        s_1+\lambda_2-\lambda_3 \geqslant \gamma\, .
    \end{cases} 
\end{equation}
This follows from the fact that $a,b,c\geqslant0$. Of course, if the momentum $\kbar_{1A'}$ is contracted with $\kbar_{3A'}$ instead of $\kbar_{2A'}$, the constraints are changed by replacing $\beta-1$ by $\beta$ and $\delta$ by $\delta-1$. The first constraint, setting a lower bound for the sum of helicities, is a consequence of the stronger condition $s_1+\lambda_2+\lambda_3=(k+\beta+\delta+2)/2$, which is just a rewriting of $\alpha+\gamma+k+1-2s_1=0$, and the fact $d$, $e$ and $f$ are positive integers.

\paragraph{$\boldsymbol{\omega \omega C}$.}

Raising once more the form degree, we now discuss amplitudes involving $\omega^{A(2s_1-2-k),A'(k)}$, $\omega^{A(2s_2-2-l),A'(l)}$ and $C^{A(\alpha),A'(\beta)}$, where $\beta-\alpha=2\lambda_3$. First, it is worth explaining the role of the differential forms in the calculation. The two gauge fields are proportional to $\ebk{q}{\kbar_1}$ and $\ebk{q}{\kbar_2}$. As the $q$'s cannot be contracted, the product of the gauge fields is proportional to $[12]\Hbk{q}{q}$.\footnote{Short for $H^{AA} q_Aq_A$.} This can only be combined with a self-dual two-form $H_{BB}$ whose indices could be contracted with any two unprimed momentum spinors except $q$. The only candidates are $k_3^B$. Thus, the amplitude is
\begin{equation}
    \mathcal{A}_3\propto \left(\frac{\ab{q3}}{\ab{q1}}\right)^{2s_1-1-k} \left(\frac{\ab{q3}}{\ab{q2}}\right)^{2s_2-1-l}  \ \rb{12}\  \rb{12}^{a} \ \rb{23}^{b} \ \rb{31}^{c} \, .
\end{equation}
The fact that the number of $q$'s must correspond to the number of $k_3$ has been used. This constraint writes $2s_1+2s_2-2-k-l=\alpha$. The factor $[12]$ comes from the product $\ebk{q}{\kbar_1}\ebk{q}{\kbar_2}$. After using momentum conservation to drop reference spinors, the only remaining ingredients in the amplitude are square brackets. Thus, even before considering special three-point kinematics (on-shell conditions), one sees that no amplitude with negative sum of helicities (with angle brackets) can be constructed. Indeed, the expression becomes
\begin{equation}
    \mathcal{A}_3\propto  \rb{12}^{a+1+2s_1+2s_2-2-k-l} \ \rb{23}^{b-2s_1+1+k} \ \rb{31}^{c-2s_2+1+l} \, .
\end{equation}
$a$, $b$ and $c$ can be determined in terms of $k$, $l$ and $\beta$. The following relations hold,
\begin{equation}
    \begin{cases}
        a=\frac{k+l-\beta}{2}\, , \\
        b=\frac{-k+l+\beta}{2}\, , \\
        c=\frac{k-l+\beta}{2}\, .
    \end{cases}\ 
    \label{systeqomomc}
\end{equation}
Of course, these numbers are all positive integers, which implies that $k$, $l$ and $\beta$ are constrained by triangular inequalities;
\begin{equation}
    \begin{cases}
        k+l\geqslant\beta\, , \\
        l+\beta\geqslant k\, , \\
        k+\beta\geqslant l\, .
    \end{cases}\ 
    \label{triklbeta}
\end{equation}
These will be expressed in terms of the helicities. With the constraints $\beta-\alpha=2\lambda_3$ and $2s_1+2s_2-2-k-l=\alpha$, the expected result is obtained:
\begin{equation}
    \mathcal{A}_3\propto \rb{12}^{s_1+s_2-\lambda_3} \ \rb{23}^{-s_1+s_2+\lambda_3}\ \rb{31}^{s_1-s_2+\lambda_3}\ .
\end{equation}
The inequalities can be massaged a little bit to show that the sum of helicities is bounded from below. Indeed,
\begin{equation}
    s_1+s_2+\lambda_3\geqslant \text{max}\{k,l,\beta\} \ ,
\end{equation}
which is not surprising since $2s_1+2s_2-2-k-l=\alpha$ implies $s_1+s_2+\lambda_3=(k+l+\beta+2)/2$. Then the last inequality is just a mere consequence of \eqref{triklbeta}.

\paragraph{$\boldsymbol{\omega \omega \omega}$.} 
The last case that must be treated is the one where all fields are one-forms. Actually, there are no non-trivial vertices of this type. It is sufficient to use the $\ebk{q}{\kbar}$-part of the gauge fields to show this result.
\begin{equation}
    \omega\wedge \omega\wedge \omega \propto e^{AA'}  \wedge e^{AB'} \wedge e^{AC'} q_A q_{A}q_{A} \kbar_{1A'}\kbar_{2B'}\kbar_{3C'} = 0 \, .
\end{equation}
The unprimed indices of the vierbein have been explicitly symmetrized. Using the identities in Appendix \ref{appdiff}, it is straightforward to see that the resulting three-form, being proportional to $e^{AA'}\wedge e^{AB'}\wedge e^{AC'}$, vanishes. Therefore, there is no cubic amplitude with more than two $\omega$'s. Moreover, for the same reason, there is no higher-point amplitude involving more than two $\omega$'s. 

\subsection{Anatomy of the interactions}

Following the discussion presented in \cite{Skvortsov:2022syz}, let us make a small comparison between the equations of motion in the light-cone gauge and the covariant formalism, making use of the FDA approach. In the former case, one can write a very schematic action\footnote{The action is a bit more intricate then the SDYM and SDGR actions, which take the form $\Lag=\Psi\Box\Phi+c_{++-}\Phi\Phi\Psi $.} 
\begin{equation}
    \Lag=\Psi\Box\Phi+c_{+++}\Phi\Phi\Phi+c_{++-}\Phi\Phi\Psi+c_{+--}\Phi\Psi\Psi \ 
\end{equation}
involving only two fields, $\Phi$ and $\Psi$, encoding, respectively, positive $\lambda\geqslant1$ and nonpositive helicities $\lambda\leqslant0$. All possible helicity structures that can appear in the amplitudes are expressed by the coefficients $c_{+\pm\pm}$. Ignoring numerical coefficients, the resulting equations of motion are
\begin{subequations}
    \begin{align}
       & \square\Phi =c_{++-}\Phi\Phi+c_{+--}\Phi\Psi\, , \label{boxphic} \\
        & \square\Psi =c_{+++}\Phi\Phi+c_{++-}\Phi\Psi+c_{+--}\Psi\Psi \, .\label{boxpsic} 
    \end{align}
\end{subequations}
In the latter case (covariant formulation), there is no complete action encoding the interactions. However, the interacting theory is known at the level of the equations of motion, which can be used to compute scattering amplitudes. The non-linear equations in the FDA approach should look as
\begin{subequations}
    \begin{align}
       & d\omega=\voo(\omega,\omega)+\voo(\omega,\omega,C)+\voo(\omega,\omega,C,C)+\ldots\, , \label{FDAom} \\
        & dC =\uoc(\omega,C)+\uoc(\omega,C,C)+\uoc(\omega,C,C,C)+\ldots \, ,\label{FDAC} 
    \end{align}
\end{subequations}
Where $\mathcal{V}$, $\mathcal{U}$ are vertices, whose explicit form will be given later. The equations can be expanded around a background value of $\omega(y,\ybar)$ denoted by $\omega_0(y,\ybar)$ and a trivial background value of $C$ ($C_0=0$). Then, the field $\omega$ can be replaced by $\omega_0+\omega$ in the equations of motion, where only quadratic terms (in the fields) are considered. Defining the background covariant derivative as $D=d+\omega_0$,\footnote{Further in the document, the operator $D$ will be denoted by the usual exterior derivative $d$.} the relevant equations are 
\begin{subequations}\label{EOMcubic}
    \begin{align}
       & D\omega =\voo(\omega,\omega)+\voo(\omega_0,\omega,C)+\voo(\omega_0,\omega_0,C,C)\, , \label{EOMcubicOm} \\
        & DC =\uoc(\omega,C)+\uoc(\omega_0,C,C) \, ,\label{EOMcubicC} 
    \end{align}
\end{subequations}
The calculations presented in the following paragraphs focus on the flat background. In this case, the value of $\omega_0(y,\ybar)$ is $\ebk{y}{\ybar}$ and will be denoted by $h(y,\ybar)$. A priori, \eqref{EOMcubicOm} and \eqref{EOMcubicC} suggest that five types of vertices should be computed to find all cubic amplitudes. Actually, only the bilinear and trilinear maps will be required. The quartic one, $\voo(h,h,C,C)$, vanishes because it contains the operator $(p_{12})^2$ acting on $h(y_1,\ybar_1)=\ebk{y_1}{\ybar_1}$ and $h(y_2,\ybar_2)=\ebk{y_2}{\ybar_2}$, which are linear in, respectively, $y_1$ and $y_2$.

As one can see by comparing the FDA equations and those obtained from the Lagrangian (in the light-cone gauge), $\omega$ and the part of $C$ that begins with $C^{A'(2s)}$ should be identified with $\Phi$, while the part of $C$ that originates at $\Psi^{A(2s)}\equiv C^{A(2s)}$ should be identified with $\Psi$. Since some of the vertices in the $\Psi$ and $\Phi$ equations come from the same Lagrangian vertex, the corresponding amplitudes must be the same. We will check this later. One complication is that $C$ contains components with both positive and negative helicity. Therefore, the helicity structure of the vertices that contain $C$ should be carefully analyzed.

\subsection{Star-product and vertex operators} \label{vertexop}

In this section, a summary of the relevant multilinear maps (vertex operators) entering equations \eqref{EOMcubicOm} and \eqref{EOMcubicC} is given. The expressions follow the conventions adopted in \cite{Skvortsov:2022syz}, or \cite{Sharapov:2022nps}. At quadratic order, the equation must satisfy the self-consistency condition implied by $d^2=0$. This enforces $\voo(\omega,\omega)$ to be an associative product (see \cite{Skvortsov:2022syz}). In terms of differential operators that we have already used when discussing the most general cubic amplitudes, the first bilinear vertex is
\begin{equation}
    \voo(\omega,\omega) = \frac{1}{2} \exp\left[p_{01} + p_{02}\right]\exp\left[\pbar_{01} + \pbar_{02} + \pbar_{12}\right] \ \omega (y_1,\ybar_1) \wedge \omega (y_2,\ybar_2)\Big|_{y_{1,2}=\ybar_{1,2}=0} \, ,
\end{equation}
where the role of the first factor is to replace $y_1$ and $y_2$ by $y$. This operation symmetrizes the unprimed indices of the fields, it is just a commutative product over $y$'s. The second factor is a Moyal-Weyl star-product, which contracts some primed indices of the fields.\footnote{The first two arguments of the second exponential can actually be ignored regarding our calculations, as the vertices that will be used are $\ybar$-independent.} The other vertices/operators will be defined without explicitly specifying the fields they act on for simplicity. The map $\uoc(\omega,C)$ actually contains two structures,
\begin{equation}
    \uocc(\omega,C) = \exp\left[p_{02}+p_{12}\right] \exp\left[\pbar_{01} + \pbar_{02} + \pbar_{12}\right] \label{u1oc}
\end{equation}
and
\begin{equation}
    \ucoc(C,\omega) = - \exp\left[p_{01}-p_{12}\right] \exp\left[\pbar_{01} + \pbar_{02} + \pbar_{12}\right]  \, . \label{u2co}
\end{equation}
Indeed, the star-product being non-commutative, the order of the arguments matters even if they are not matrix-valued. Chiral Theory can be enriched by adding a matrix structure. It admits $U(N)$, $O(N)$ and $USp(N)$ gaugings.\footnote{The full theory contains interactions even without gaugings of Yang--Mills type. For instance, interactions in the pure gravity sector can be defined without any matrix-valued fields. However, such a structure is required when considering interactions in Higher-Spin Self-Dual Yang--Mills (HS-SDYM), which will be used in a later section.} The last objects we need to compute amplitudes are the $\voo$- and $\uoc$-trilinear maps.\footnote{The expressions are borrowed from \cite{Skvortsov:2022syz} and \cite{Sharapov:2022nps}.} The former have the following expressions
\begin{equation}
\begin{aligned}
    \vooc(\omega,\omega,C)&=p_{12}\ S\int_{\Delta_2}\exp \left[(1-u) p_{01}+(1-v) p_{02}+up_{13}+v p_{23}\right] \, , \\
    \voco(\omega,C,\omega)&=-p_{13}\ S\int_{\Delta_2} \exp \left[(1-u) p_{01}+(1-v) p_{03}+u p_{12}-v p_{23}\right] \\
    &\ \ \  -p_{13}\ S\int_{\Delta_2} \exp \left[(1-v) p_{01}+(1-u) p_{03}+v p_{12}-u p_{23}\right]\, , \\
    \vcoo(C,\omega,\omega)&=p_{23}\ S\int_{\Delta_2} \exp \left[(1-v) p_{02}+(1-u) p_{03}-v p_{12}-u p_{13}\right] \, , 
\end{aligned}
\label{triliV}
\end{equation}
and the latter are defined as
\begin{equation}
\begin{aligned}
\mathcal{U}_{1}(\omega,C,C)= & \ +p_{01}\ S \int_{\Delta_{2}} \exp \left[(1-v) p_{02}+v p_{03}+(1-u)p_{12}+u p_{13}\right]\ , \\
\mathcal{U}_{2}(C,\omega,C)= & -p_{02}\ S \int_{\Delta_{2}} \exp \left[v p_{01}+(1-v) p_{03}-u p_{12}+(1-u) p_{23}\right] \\
& -p_{02}\ S \int_{\Delta_{2}} \exp \left[u p_{01}+(1-u) p_{03}-v p_{12}+(1-v) p_{23}\right]  \ ,\\
\mathcal{U}_{3}(C,\omega,\omega)= & +p_{03}\ S \int_{\Delta_{2}} \exp \left[(1-u) p_{01}+u p_{02}-(1-v) p_{13}-v p_{23}\right] \ .
\end{aligned}
\label{triliU}
\end{equation}
The operator $S=\exp\left[\pbar_{01}+\pbar_{02}+\pbar_{03}+\pbar_{12}+\pbar_{13}+\pbar_{23}\right]$ represents the Moyal-Weyl product of three functions. Higher-order operators, such as $\voo(\omega,\omega,C,C)$, which appears in \eqref{EOMcubicOm}, are defined in \cite{Sharapov:2022nps} and \cite{Sharapov:2023erv}. 

\subsection{Three-point amplitudes from the equations of motion}

When computing amplitudes, it is useful to keep track of the helicities. In this purpose, some extra parameters can be inserted into the free fields. As such,
\begin{align}
    \omega(x|y,\bry)&=\frac{\ebk{q}{\kbar}}{\ab{qk}+\sigma yq}e^{\sigma\bry \brk}\, , & C(x|y,\bry)=\tfrac{1}{2}\exp(\sigma yk+\sigma^{-1}\bry\bar{k}) \, ,
    \label{solgenfnhel}
\end{align}
where the parameter $\sigma$ is defined such that, when projecting the generating function $\omega(x|y,\bry)$ onto a component of spin $s$, a factor $\sigma^{2s-2}$ is produced. In the case of $C(x|y,\bry)$, the factor is $\sigma^{-2\lambda}$. This is due to the fact that, concerning the zero-form field, the difference between the number of primed indices and the number of unprimed indices gives twice the helicity.
The procedure to compute amplitudes is the following. Starting from the free action 
\begin{equation}
    S=\sum_{s=1}^\infty\int\Psi^{A(2s)}H_{AA}\wedge d\omega_{A(2s-2)}\, ,
    \label{freeactionsum}
\end{equation}
the derivatives $d\omega$ and $d\Psi$ ($=dC(\ybar=0)$) can be alternatively replaced by the right-hand sides of the non-linear equations \eqref{EOMcubic} to construct cubic amplitudes.\footnote{As a rule, the equations \eqref{FDAom} and \eqref{FDAC} can be used to compute amplitudes order by order, but explicit calculations are carried out only for the cubic sector further in text.} A remark must be made about the vertex operators that have been listed previously. Since the physical fields $\Psi^{A(2s)}$ and $\omega_{A(2s-2)}$ that appear in the action \eqref{freeactionsum} contain only unprimed indices, the $\ybar$-dependence can be dropped in all the vertices $\voo(\omega,\omega,\ldots)$ and $\uoc(\omega,C,\ldots)$. For example, in terms of generating functions, a cubic amplitude can be constructed from $\voo(\omega,\omega)$ via the expression
\begin{equation}
    \ampl_3=\exp\left(\pl_1\cdot\pl\right) \int C(y_1) \Hbk{y}{y} \voo(\omega,\omega)(y)\Big|_{y_1=y=0} \, ,
    \label{amplfromVU}
\end{equation}
where the fields are the free solutions \eqref{solgenfnhel}. The result is thus a generating function containing (infinitely-) many amplitudes, in the sense that it involves several different combinations of spins. Then, the parameters $\sigma_i$ (denoting the helicity of field number $i$) allow extracting specific combinations of helicities. The procedure can be summarized in three steps: (a) Apply a vertex operator from Section \ref{vertexop} to the free solutions, (b)  Contract the result with a physical field, (c) Project the resulting amplitude onto a triplet of fixed helicities.

Following the discussion in \ref{vertexop}, four different structures can be used to extract cubic amplitudes, namely $\voo(\omega,\omega)$, $\voo(h,\omega,C)$, $\uoc(\omega,C)$ and $\uoc(h,C,C)$. 

\paragraph{$\boldsymbol{\voo(\omega,\omega)}$.} Let us start with the simplest map that appears in the non-linear equation of the gauge field. Applying it to free field solutions, the following result is obtained.
\begin{align*}
    \voo(\omega,\omega)&= e^{p_{01}+p_{02}+\pbar_{12}}\ \omega(y_1,\ybar_1)\wedge\omega(y_2,\ybar_2)\\
    &=e^{AA'}\wedge e^{BB'}\ q_A \kbar_{1A'}q_B\kbar_{2B'}\ e^{\pbar_{12}}\ \frac{e^{\sigma_1\ybar_1\kbar_1}}{\ub{q1}+\sigma_1yq}\frac{e^{\sigma_2\ybar_2\kbar_2}}{\ub{q2}+\sigma_2yq}\ e^{\pm k_1x\kbar_1\pm k_2x\kbar_2}\\
    &=-\frac{1}{2}\ \Hbk{q}{q}\ [12]\ e^{\sigma_1\sigma_2[12]}\ \frac{1}{\ub{q1}+\sigma_1yq}\frac{1}{\ub{q2}+\sigma_2yq}\ e^{\pm k_1x\kbar_1\pm k_2x\kbar_2} \, .
\end{align*}
Of course, it is understood that the variables $y_{1,2}$ and $\ybar_{1,2}$ are set to zero after letting the associated differential operators act on them. 
Shifting the variables $1\rightarrow2$ and $2\rightarrow3$ of $\voo(\omega,\omega)$, the amplitude can be computed using \eqref{amplfromVU}, as well as identities from Appendix \ref{appdiff}, and integrating out the space-time dependence. The result is 
\begin{equation}
    \ampl_3=-\frac{1}{12}\ \sigma_1^2\ [23]\ \frac{\ub{q1}}{\ub{q2}-\sigma_1\sigma_2\ub{q1}}\frac{\ub{q1}}{\ub{q3}-\sigma_1\sigma_3\ub{q1}} e^{\sigma_2\sigma_3[23]} \, .
\end{equation}
The consequences of the momentum conservation \eqref{momconsqk} allow us to rewrite the expression as
\begin{equation}
    \ampl_3=-\frac{1}{12}\  \sigma_1^2 [23]\  \frac{[23]}{[31]-\sigma_1\sigma_2[23]}\frac{[23]}{[12]-\sigma_1\sigma_3[23]} e^{\sigma_2\sigma_3[23]} \, ,
\end{equation}
thereby cancelling all the reference spinors. The last step consists of performing a Taylor expansion of the amplitude, written as a generating function, to obtain a sum of helicity-amplitudes that exhibit the expected structure \eqref{genericV}.
\begin{equation}
    \ampl_3=-\frac{1}{12}\sum_{l,n,m}\frac{1}{l!} \ \sigma_1^{n+m+2}\sigma_2^{n+l}\sigma_3^{m+l} \ [23]^{n+m+2}[12]^{-m-1}[31]^{-n-1} \, ,
\end{equation}
where the sums are, for now, independent of one another and run from zero to infinity. Using the fact that the powers of the parameters $\sigma_i$ are related to the helicity of each corresponding field, the following constraints can be imposed.
\begin{equation}
    \begin{cases}
        & \hspace{-4mm} n+m+2=-2\lambda_1 \, ,\\
        & \hspace{-4mm} n+l=2s_2-2 \, ,\\
        & \hspace{-4mm} n+l=2s_3-2 \, .
    \end{cases}
\end{equation}
This system of equations allows expressing the integers $n$, $m$ and $l$ in terms of the helicities. For example, we have $l+1=\lambda_1+s_2+s_3$, which yields the expected coefficient $1/\Gamma$ appearing in \eqref{eq:magicalcoupling}. Notice that $\lambda_1=-s_1$, as the integers $n$ and $m$ are non-negative. The reason is that the field $C(y_1)$ involved in the amplitude is a physical field with negative helicity. For a specific triplet of spins, we have the following helicity-amplitude
\begin{equation}
     \ampl_3^{-s_1,+s_2,+s_3}=-\frac{1}{12} \ \frac{1}{\Gamma(-s_1+s_2+s_3)}\ [12]^{-s_1+s_2-s_3}[23]^{s_1+s_2+s_3}[31]^{-s_1-s_2+s_3}\, ,
\end{equation}
which matches the well-known result \eqref{genericV} that was originally obtained in the light-cone gauge. This amplitude was reproduced in \cite{Skvortsov:2022syz}.

\paragraph{$\boldsymbol{\voo(h,\omega,C)}$.}As the theory can be expanded around the background solution ($h=\ebk{y_1}{\ybar_1}$, $C_0=0$), some cubic interactions (quadratic from the equations of motion point view) can be extracted from trilinear maps $\voo(\omega,\omega,C)$ and $\uoc(\omega,C,C)$. At quadratic order, the equation of motion can be written as
\begin{equation}
    d\omega =\voo(\omega,\omega)+\woo(\omega,C)+\woo(C,\omega) \, ,  \label{EOMomWoc}
\end{equation}
where the term $\voo(h,\omega,C)$ has been divided into two contributions, $\woo(\omega,C)$ and $\woo(C,\omega)$, according to the ordering of the fields. Their respective expressions are 
\begin{equation}
    \woo(\omega,C)=\vooc(h,\omega,C)+\vooc(\omega,h,C)+\voco(\omega,C,h) \, , \label{wocV}
\end{equation}
and 
\begin{equation}
    \woo(C,\omega)=\voco(h,C,\omega)+\vcoo(C,\omega,h)+\vcoo(C,h,\omega) \, . \label{wcoV}
\end{equation}
The expressions \eqref{solgenfnhel} are used for the calculation of both contributions. Although they render the computations quite unaesthetic, the $\sigma$'s will be required to project the rather abstract generating functions onto amplitudes with specific helicity structures. Associating labels 2 and 3 to both momenta and twistor variables of $\omega$ and $C$, respectively, \eqref{wocV} takes the form
\begin{align*}
    \woo(\omega,C) &=-\frac{1}{2}\ \Hbk{q}{q}\ \sigma_2\sigma_3^{-1}\pb{23}\ e^{\sigma_2\sigma_3^{-1}\pb{23}}\\
    &\ \ \ \int_{\Delta_2} \left(e^{(1-u)p_{02}+up_{23}}+e^{(1-v)p_{02}+vp_{23}}\right) \ \frac{1}{(\ub{q2}+\sigma_2y_2q)^2} \ e^{\sigma_3y_3k_3} \, .
\end{align*}
The expression that must be integrated over the simplex $\Delta_2=\{0\leqslant u<v\leqslant 1\}$ is
\begin{equation}
    \frac{1}{(\ub{q2}+\sigma_2[(1-u)yq+u\sigma_3\ub{q3}])^2}+\frac{1}{(\ub{q2}+\sigma_2[(1-v)yq+v\sigma_3\ub{q3}])^2} \ .
\end{equation}
Using some condensed notations to compute the integral, one finds
\begin{equation}
    \int_{\Delta_2} \left(\frac{1}{(A+Bu)^2}+\frac{1}{(A+Bv)^2}\right)=\frac{1}{A(A+B)} \, .
\end{equation}
Therefore, the second term on the right-hand side of \eqref{EOMomWoc} takes the following form
\begin{equation}
    \woo(\omega,C) =-\frac{1}{2}\ \Hbk{q}{q}\ \sigma_2\sigma_3^{-1}\pb{23}\ e^{\sigma_2\sigma_3^{-1}\pb{23}}\ \frac{1}{\ub{q2}+\sigma_2yq}\frac{1}{\ub{q3}+\sigma_2\sigma_3\ub{q3}}\ e^{\pm k_2x\kbar_2\pm k_3x\kbar_3} \, .
    \label{wocfinal}
\end{equation}
The remaining term of \eqref{EOMomWoc} takes a very similar form
\begin{equation}
    \woo(\omega,C) =\frac{1}{2}\ \Hbk{q}{q}\ \sigma_2\sigma_3^{-1}\pb{23}\ e^{-\sigma_2\sigma_3^{-1}\pb{23}}\ \frac{1}{\ub{q2}+\sigma_2yq}\frac{1}{\ub{q3}+\sigma_2\sigma_3\ub{q3}}\ e^{\pm k_2x\kbar_2\pm k_3x\kbar_3} \, .
    \label{wcofinal}
\end{equation}
Remark: The two expressions differ only by a minus sign and by a sign in the exponential. A consequence of this is that, for odd spins, the amplitudes computed from the sum of \eqref{wocfinal} and \eqref{wcofinal} vanish. This is true unless color is added into the theory by letting the fields take values in a Lie algebra. Contracting both expressions with $C(y_1,\ybar_1)$ and applying the usual techniques to both expressions yields 
\begin{equation}
    \ampl_3^{-s_1,+s_2,\mp s_3}= -\frac{1}{12}\ \frac{(-1)^{-s_1+s_2\pm s_3}}{\Gamma(-s_1+s_2\mp s_3)} \pb{12}^{-s_1+s_2\pm s_3} \pb{23}^{s_1+s_2\mp s_3} \pb{31}^{-s_1-s_2\mp s_3}
\end{equation}
and 
\begin{equation}
    \ampl_3^{-s_1,+s_2,\mp s_3}= -\frac{1}{12}\ \frac{1}{\Gamma(-s_1+s_2\mp s_3)} \pb{12}^{-s_1+s_2\pm s_3} \pb{23}^{s_1+s_2\mp s_3} \pb{31}^{-s_1-s_2\mp s_3} \, .
\end{equation}

\paragraph{$\boldsymbol{\uoc(\omega,C)}$.}To derive amplitudes via the second equation of \eqref{EOMcubic}, one can consider the free action and perform an integration by parts. Let us compute the amplitude associated to $\uoc_1(\omega,C)$ by using 
\begin{equation}
    \ampl_3=\exp\left(\pl_1\cdot\pl\right) \int \omega(y_1) \Hbk{y_1}{y_1}\ \uocc(\omega,C)(y)\Big|_{y_1=y=0} \, ,
\end{equation}
and following the same steps as for $\voo(\omega,\omega)$. The expression of the amplitude is, after making use of momentum conservation,
\begin{equation}
\ampl_3=-\frac{1}{12}\  \sigma_3^2\ [12]\  \frac{[12]}{[23]+\sigma_1\sigma_3[12]}\frac{[12]}{[31]+\sigma_2\sigma_3[12]}\ e^{\sigma_2\sigma_3^{-1}[23]}\, ,
\end{equation}
Expanding the generating functions leads to
$$
\ampl_3=-\frac{1}{12}\ \sum_{l,n,m} \frac{(-1)^{n+m}}{l!}\ \sigma_1^{n}\sigma_2^{m+l}\sigma_3^{n+m-l+2} \ \pb{12}^{n+m+3}\pb{23}^{-n-1+l}\pb{31}^{-m-1} \, .
$$
This time, the field $C$ can potentially carry both helicities, as it is not required to be a physical field. The result is
\begin{equation}
    \ampl_3^{+s_1,+s_2,\mp s_3} = \frac{1}{12} \ \frac{(-1)^{s_1-s_2\pm s_3}}{\Gamma(s_1+s_2\mp s_3)} \ \pb{12}^{s_1+s_2\pm s_3}\pb{23}^{-s_1+s_2\mp s_3}\pb{31}^{s_1-s_2\pm s_3} \, .
\end{equation}
The extra sign can be absorbed by any of the bracket products. In all cases, the amplitude has the expected form. This amplitude was found in \cite{Skvortsov:2022syz}.

\paragraph{$\boldsymbol{\uoc(h,C,C)}$.} The second equation of motion \eqref{EOMcubic} can be rewritten as follows.
\begin{equation}
    dC  =\uoc(\omega,C)+\xcc(C,C) \, ,
\end{equation}
where 
\begin{equation}
    \xcc(C,C)=\uocc(h,C,C)+\ucoc(C,h,C)+\ucco(C,C,h)\,.
\end{equation}
Let us evaluate the action of $\uocc$ when $\omega$ is replaced by the background field $h(y,\ybar)=\ebk{y}{\ybar}$.
\begin{equation*}
    \begin{aligned}
\uoc_{1}(h, C, C) & = p_{01} \int_{\Delta_{2}} e^{(1-u) p_{02}+u p_{03}}\left(\bar{p}_{12}+\bar{p}_{13}\right) e^{\bar{p}_{23}} \ h(y_1,\ybar_1)\ C(y_2,\ybar_2)\ C(y_3,\ybar_3) \\
& =\int_{\Delta_{2}} f(u)\left(\bar{p}_{12}+\bar{p}_{13}\right) e^{\bar{p}_{23}}\ \ebk{y}{\ybar_{1}} \ C(y_2,\ybar_2)\ C(y_3,\ybar_3) \\
& =\int_{\Delta_{2}} f(u) e^{(\sigma_{2}\sigma_{3})^{-1}[23]}\ \left[-\sigma_{2}^{-1}\ebk{y}{\bar{k}_{2}}-\sigma_{3}^{-1}\ebk{y}{\bar{k}_{3}} \right] \ C(y_2,\ybar_2)\ C(y_3,\ybar_3) \, ,
\end{aligned}
\end{equation*}
where $\exp \left[(1-u) p_{02}+u p_{03}\right]$ has been denoted by $f(u)$. We used $$p_{01}\bar{p}_{1i}\ h(y_1,\ybar_1)C(y_i,\ybar_i)=-\sigma_i^{-1} \ebk{y}{\bar{k}_{i}}C(y_i,\ybar_i)\,, \qquad \quad i\in \{2,3\}\,.$$ Before computing the amplitude, let us apply the relation 
$$
\pbar_{1i}\ \omega(y_1)\Hbk{y_1}{y_1}\ h(y_1,\ybar_1)C(y_i,\ybar_i)=\frac{1}{6}\ \sigma_i^{-1}\pb{1i}\frac{\ub{y_1y}\ub{y_1q}}{\ub{q1}+\sigma_1y_1q} C(i)\  H\wedge H \, .
$$
Then,
$$
\begin{aligned}
    \ampl_3
    &= \frac{1}{6} \ e^{\pl_1\cdot\pl} \int_{\Delta_{2}} f(u)  (\sigma_{2}\pb{12}+\sigma_{3}\pb{13}) e^{(\sigma_{2}\sigma_{3})^{-1}[23]}\ C(2) C(3)\\
    &=\frac{1}{24} \ e^{(\sigma_{2}\sigma_{3})^{-1}[23]}  (\sigma_{2}^{-1}\pb{12}+\sigma_{3}^{-1}\pb{13}) \int_{\Delta_{2}}  \left(\frac{3\alpha(u)}{[23]+\sigma_1\alpha(u)}-\sigma_1 \left(\frac{\alpha(u)}{[23]+\sigma_1\alpha(u)}\right)^2\right)\, ,
\end{aligned}
$$
where $\alpha(u)=\sigma_3(1-u)\ub{q3}+\sigma_2u\ub{q2}$. It is rather trivial to see that the amplitude involves gauge-invariant ratios, where $\ub{q1}$, $\ub{q2}$ and $\ub{q3}$ can be respectively replaced by $[23]$, $[31]$ and $[12]$. As a result, $\alpha(u)$ can be redefined as $\alpha(u)=\sigma_3(1-u)[12]+\sigma_2u[31]$ Let us denote the integrand of the amplitude by $F(u)$. Following the same steps for $\ucoc$ and $\ucco$ yields a rather cumbersome expression for the total amplitude.
\begin{equation}
\begin{aligned}
    \ampl_3 & = \frac{1}{24} \ e^{(\sigma_{2}\sigma_{3})^{-1}[23]} \int_{\Delta_{2}} [ (\sigma_{2}^{-1}[12]-\sigma_{3}^{-1}[31]) (F(v)-F(u))\\
    & \ \ \ \ \ \ \ \ \ \ \ \ \ \ \ \ \ \ \ \ \ \ \ \  +(\sigma_{2}^{-1}[12]+\sigma_{3}^{-1}[31])(F(1-v)+F(1-u))] 
\end{aligned}
    \label{totalamplitudeU}
\end{equation}
Let us perform a Taylor expansion in $\sigma_1$ to check that the previous expression contains helicity-amplitudes with helicity-dependent coefficient equal to $[\Gamma(\lbd_1+\lbd_2+\lbd_3)]^{-1}$. 
Given the results derived above, we have all the necessary ingredients for the last steps of the calculation. After performing the integral, which is done in Appendix C, the total cubic amplitude writes
\begin{equation}
\begin{aligned}
    \ampl_3 = \frac{1}{12} \ e^{(\sigma_{2}\sigma_{3})^{-1}[23]} &\left(\sum_{s_1=1}^\infty [23]^{1-2s_1}\sum_{n=0}^{2s_1-1} \frac{(n+1)}{2s_1}\ [12]^{2s_1-1-n}[31]^{n+1}\sigma_2^{n}\sigma_3^{2s_1-2-n}\right. \\ 
    & \ \ \ +\left.\sum_{s_1=1}^\infty [23]^{1-2s_1}\sum_{m=0}^{2s_1-1} \frac{(2s_1-m)}{2s_1}\ [12]^{2s_1-m}[31]^{m}\sigma_2^{m-1}\sigma_3^{2s_1-1-m}\right) \, . \\ 
\end{aligned}
\label{ampltriliUexpanded}
\end{equation}
As usual, let us expand the exponential as $\sum_{l=0}^\infty (1/l!)\ \sigma_2^{-l}\sigma_3^{-l}[23]^l$, and relate the powers of the $\sigma$'s to the helicities. For the first term, we have
\begin{equation}
    \left\{\begin{aligned}
       &n-l =-2\lbd_2\, , \\
       &2s_1-2-n-l = -2\lbd_3\, .
    \end{aligned}
    \right.
\end{equation}
Then,
\begin{equation}
    \left\{\begin{aligned}
       &l+1 =s_1+\lbd_2+\lbd_3\, , \\
       &n = s_1-\lbd_2+\lbd_3-1\, .
    \end{aligned}
    \right.
\end{equation}
For the second term, we have
\begin{equation}
    \left\{\begin{aligned}
       &m-1-l =-2\lbd_2\, , \\
       &2s_1-1-m-l = -2\lbd_3\, .
    \end{aligned}
    \right.
\end{equation}
Hence, one finds
\begin{equation}
    \left\{\begin{aligned}
       &l+1 =s_1+\lbd_2+\lbd_3\, , \\
       &m = s_1-\lbd_2+\lbd_3-1\, .
    \end{aligned}
    \right.
\end{equation}
Forgetting about the summation over the helicities, the cubic amplitude now becomes
$$
\begin{aligned}
    \ampl_3&=\frac{1}{12} \frac{1}{\Gamma(s_1+\lbd_2+\lbd_3)} \ [12]^{s_1+\lbd_2-\lbd_3}\ [23]^{-s_1+\lbd_2+\lbd_3}\ [31]^{s_1-\lbd_2+\lbd_3}\ \frac{s_1+\lbd_2-\lbd_3}{2s_1} \\
    & \ \ \ + \frac{1}{12} \frac{1}{\Gamma(s_1+\lbd_2+\lbd_3)}\ [12]^{s_1+\lbd_2-\lbd_3}\ [23]^{-s_1+\lbd_2+\lbd_3}\ [31]^{s_1-\lbd_2+\lbd_3}\ \frac{s_1-\lbd_2+\lbd_3}{2s_1} \, .
\end{aligned}
$$
Finally, 
\begin{equation}
    \ampl_3=\frac{1}{12} \frac{1}{\Gamma(s_1+\lbd_2+\lbd_3)} \ [12]^{s_1+\lbd_2-\lbd_3}\ [23]^{-s_1+\lbd_2+\lbd_3}\ [31]^{s_1-\lbd_2+\lbd_3}
\end{equation}
To conclude, we reproduced the expected cubic amplitudes from the Lorentz-covariant equations of motion.

\section{A word about quartic and higher contact terms} \label{potquartic}

The infinite set of multilinear maps on the right-hand side of equations \eqref{FDAom} and \eqref{FDAC}, expressing the theory as an FDA, is required for Lorentz covariance of the theory. However, it is possible to show that all tree-level quartic amplitudes constructed from $\voo(\omega,\omega,C)$ and $\uoc(\omega,C,C)$ vanish. Using the plane-wave solutions as well as very schematic expressions for the vertex operators leads straightforwardly to the result. Let us start with the expressions
\begin{equation}
    \voo(\omega,\omega,C)=p_{23}\int_{\Delta_2} \sum_i \mathcal{O}_i(y,u,v)\  \omega(y_2,\ybar_2)\ \omega(y_3,\ybar_3)\ C(y_4,\ybar_4)
\end{equation}
and 
\begin{equation}
    \uoc(\omega,C,C)=p_{02}\int_{\Delta_2} \sum_i \mathcal{O}_i(y,u,v)\  \omega(y_2,\ybar_2)\ C(y_3,\ybar_3)\ C(y_4,\ybar_4)\, ,
\end{equation}
where the way has been paved to compute the amplitude by introducing a physical field $\omega(y_1,\ybar_1)$ or $C(y_1,\ybar_1)$. $\mathcal{O}_i(y,u,v)$ are operators depending on the variables $u$ and $v$ as well as the twistor variables $y^A$. One directly sees that the former expression is zero if the fields are replaced by the plane-wave solutions. Indeed, the operator $p_{23}$ contracts two auxiliary momentum spinors $q_A$ together. This, of course, requires to choose one and the same reference spinor for all positive helicity plane-waves. The situation is slightly less trivial for the latter expression, but the same conclusion can be drawn by examining the corresponding amplitude. 
$$
\begin{aligned}
    \ampl_4
    &= e^{\pl\cdot\pl_1}\int_M \omega(y_1,\ybar_1)\Hbk{y_1}{y_1}\ p_{02}\int_{\Delta_2} \sum_i \mathcal{O}_i(y,u,v)\  \omega(y_2,\ybar_2)\ \omega(y_3,\ybar_3)\ C(y_4,\ybar_4) \\
    &= -p_{12}\int_{\Delta_2} \sum_i \mathcal{O}_i(-\pl_1,u,v)\int_M \ebk{q}{\kbar_1}\ebk{q}{\kbar_2}\Hbk{y_1}{y_1} \  f_1(y_1q)f_2(y_2q)\ C(y_3,\ybar_3)\ C(y_4,\ybar_4) \, .
\end{aligned}
$$
For convenience, $\omega$ has been expressed as $\ebk{q}{\kbar}f$, where $f$ encodes the dependence on both twistor variables and space-time coordinates. Using the identities presented in Appendix \ref{appdiff} is sufficient to determine that the amplitude is trivial. As a matter of fact,
$$
p_{12}\ \ebk{q}{\kbar_1}\ebk{q}{\kbar_2}\Hbk{y_1}{y_1} \  f_1(y_1q)\ f_2(y_2q)
     \propto [12]\ p_{12} \ (y_1q)^2\  f_1(y_1q)\ f_2(y_2q) = 0 \, .
$$
Once again, the operator $p_{12}$ contracts a pair of $q_A$'s.

Let us make a few remarks. We took the case of where the ordering of the fields does not matter (no gauging). This assumption was certainly not necessary to show that the $\voo$-contribution vanishes. The conclusion here is that higher-order maps do not contribute to the amplitudes. Indeed, the arguments can easily be extended to higher order vertices involving more $C$ fields. In the next section, we develop some recursive machinery to prove that, at least for a specific (but nonetheless interesting) truncation of Chiral HiSGRA, all amplitudes are trivial beyond the cubic sector.

\section{Berends--Giele currents in HS-SDYM} \label{BGsec}

Recursive methods are a very efficient tool for constructing amplitudes. One can cite the well-known BCFW recursion, originally conjectured by Britto, Cachazo, and Feng in \cite{BCF}, afterwards proven in \cite{BCFW}. This technique is a particular case of \textit{on-shell} recursion. In this section, our interest is devoted to another, \textit{off-shell}, recursive technique called the Berends--Giele recursion \cite{Berends:1987me}. A brief review of this tool is given in Appendix \ref{appBGC}, as well as a detailed construction of currents in a simple spin-1 theory (SDYM). Some interesting publications that include a discussion on currents in SDYM are \cite{Krasnov:2016emc}, \cite{chattopadhyay2023aspectsSDYM}, and \cite{Monteiro_2011}. The first paper discusses currents in both self-dual Yang--Mills (SDYM) and self-dual Gravity (SDGR). The second source includes a detailed derivation of the SDYM current for any number of on-shell legs.\footnote{The main difference with the one presented in Appendix \ref{appBGC} is that our proof includes generating functions, which are most convenient in order to generalize the results to arbitrary spin.} Long story short, in full Yang--Mills,\footnote{The purpose of this example is to present a familiar formula with vector indices. The SDYM currents are commonly expressed with a pair of spinor indices, whereas higher-spin currents involve many indices. This motivates the use of generating functions.} an amplitude can be computed via the following relation
\begin{equation}
    \ampl_n=\eps_\mu J^\mu_{n-1}\, , \label{amplJeps}
\end{equation}
where $\eps^\mu$ is a polarization vector, and $J^\mu_{n-1}$ is, roughly speaking, ``almost'' an amplitude with $n-1$ on-shell legs and a single off-shell leg. Contracting this object with a polarization vector while taking the on-shell limit produces an fully on-shell physical amplitude of order $n$.

Here, the aim is to construct Berends--Giele currents of arbitrary order in the case of Higher-Spin Self-Dual Yang--Mills theory, which is a consistent subsector of Chiral HiSGRA \cite{Ponomarev:2017nrr,Krasnov:2021nsq}, see also \cite{Monteiro:2022xwq,Serrani:2025owx} for a more complete picture, and subsequently to show that all amplitudes vanish beyond the cubic sector. The theory has the advantage of including infinitely-many interactions of higher-spin fields, in the sense that all integer spins are contained in the theory. However, only a subset (but still infinite) of the coupling constants $C_{\lbd_1,\lbd_2,\lbd_3}$ mentioned in Section \ref{sectionAmpl} are non-vanishing. They satisfy the constraint $\lbd_1+\lbd_2+\lbd_3=1$. All vertices are cubic and have a simple SDYM-like structure. The HS-SDYM action can be written as follows \cite{Krasnov:2021nsq}:
\begin{equation}
    S=\sum_{s=1}^\infty\ \trace\int \Psi^{A(2s)}H_{AA} \wedge\left(d\omega_{A(2s-2)}-\sum_{s_1+s_2=s+1}\omega_{A(2s_1-2)}\wedge\omega_{A(2s_2-2)} \right) \, .
    \label{actionHSSDYM}
\end{equation}
This action contains only zero-derivative vertices and all fields are physical. By contrast, if one manages to construct a complete action for Chiral Theory, the interactions terms would contain vertices with many derivatives which would be encoded in auxiliary fields such as $\omega_{A(2s-2-k),A'(k)}$ and $ \Psi^{A(2s+l),A'(l)}$ ($k\neq 0$, $l\neq 0$). Of course, all fields take values in the adjoint representation of a compact Lie algebra, even though no color coefficient will appear explicitly in this section, since the currents are color-stripped objects.

The main ingredients needed to apply the Berends--Giele recursion are the free propagator and the elementary currents (= polarization tensors).\footnote{\cite{Monteiro_2011} contains detailed calculations for the scalar field and SDYM formulated in the light-cone gauge. The reader can refer to this source for more justifications of the perturbative method used in this section.} These objects shall be written in the language of generating functions for the sake of notational simplicity. First, let us write the equation of motion in terms of zero-form gauge fields. Varying the action \eqref{actionHSSDYM} with respect to $\Psi$ leads to the equation of motion
\begin{equation}
   (y\, H\,y)\wedge [ d\omega(y)-\omega(y)\wedge \omega(y)]=0 \ ,
\end{equation}
This notation is indeed more aesthetic than $H_{AA}\wedge (d\omega_{A(2s-2)}-\sum_{s_1+s_2=s+1}\omega_{A(2s_1-2)}\wedge\omega_{A(2s_2-2)})$. Defining $\omega(y)=\ebk{\pl}{\plbar}\varphi(y,\ybar)$, where $\varphi(y,\ybar)$ is linear in $\ybar$, and projecting the equation onto its self-dual components (see Appendix \ref{appdiff}), the equation becomes 
\begin{equation}
    \pl_A \pl_{AA'}\varphi(y)^{A'}=\pl_A\varphi(y)_{A'} \pl_A\varphi(y)^{A'}\,.
\end{equation}
It is convenient to contract the equation with $y^Ay^A$ and define the number (or Euler) operator as $N\equiv y^A\pl_A$. The notation is justified by the fact the Euler operator counts the power of $y^A$ in each monomial of a generating function. We obtain
\begin{equation}
    y^A \pl_{AA'}(N\varphi(y)^{A'})=(N\varphi(y)_{A'}) (N\varphi(y)^{A'}) \, .
\end{equation}
The last manipulation consists of rescaling the fields $\phi=N\varphi$. The final form of the non-linear equation of motion is thus
\begin{equation}
    y^A \pl_{AA'}\phi(y)^{A'}=\phi(y)_{A'}\phi(y)^{A'} \, .
    \label{EOMHSSDYM}
\end{equation}
Knowing the expression of the free gauge field $\omega$ \eqref{solgenfn}, the expression of the new field $\phi$ can be determined. 
\begin{equation}
    \phi(y)^{B'}=\frac{yq}{\ub{qk}+yq}\kbar^{B'} e^{kx\kbar} \, .
\end{equation}
From this expression, the elementary current can be deduced
\begin{equation}
    \mathcal{J}^{(0)}(y)^{B'}=\frac{yq}{\ub{qk}+yq}\kbar^{B'} \, .
    \label{phi0HS}
\end{equation}
This object carries polarization tensors for all (integer) spins. The superscript "$(0)$" emphasizes that $\mathcal{J}^{(0)}$ is the zeroth-order current.\footnote{This convention is borrowed from \cite{Monteiro_2011}.} Higher-order currents will be denoted by $\mathcal{J}_{12\ldots n}^{(n-1)}$, where the superscript and subscript denote the order (in the coupling constant) and the number on-shell legs, respectively. For more information on Berends--Giele currents, the reader can refer to Appendix \ref{appBGC} where the spin-1 case is reviewed in detail.

The two-point current, or equivalently, the first-order current $\mathcal{J}^{(1)}$ is constructed by merging two functions $\mathcal{J}^{(0)}_1$ and $\mathcal{J}^{(0)}_2$ - depending on momenta $k_1$ and $k_2$ - with the free propagator of the theory. Let us define it as an operator acting on generating functions, in momentum space. It is not hard to guess that the expression we are after is proportional to $\frac{1}{k^2}(\pl_Bk^{BB'})$. Indeed, by definition, the propagator is the inverse of the kinetic operator in \eqref{EOMHSSDYM}.\footnote{In momentum space, the derivative $\pl_{CC'}$ is replaced by $k_{CC'}$.} A few lines of calculations yield the following result:
\begin{equation}
    \frac{1}{k^2}(\pl_Bk^{BB'})(y^Ck_{CC'}\phi(y)^{C'})=(N+1) \phi(y)^{B'} \, ,
\end{equation}
where the right-hand side of the equality is obtained by using the identities
$T_{AB}=T_{BA}+\eps_{AB}\ T_{C}{}^C$ and $k_{CC'}k^{CB'}=k^2\eps_{B'}{}^{C'}$, as well as the Lorenz condition $\pl^Ck_{CC'}\phi(y)^{C'}=0$. The propagator that acts on generating functions is thus defined as\footnote{More systematically, one can add the standard gauge-fixing term for the Lorenz gauge. As explained in \cite{Krasnov:2021nsq}, the Nakanishi-Lautrup field can be combined with $\Psi^{A(2s)}$ to form a ``traceful'' $\Psi^{A(2s-1);B}$. However, the vertices contain only the symmetric part of $\Psi^{A(2s-1);B}$, which is $\Psi^{A(2s)}$ itself. Therefore, we can just symmetrize the indices to begin with. This corresponds to the strict Lorenz gauge, i.e. when $\pl_{BB'}\Phi^{A(2s-2)B,B'}=0$. }
\begin{equation}
    \Delta(k)\equiv \frac{1}{N+1}(\pl_B\pi^{BB'})\label{genpropag} \, ,
\end{equation}
where $\pi^{BB'}\equiv k^{BB'}/k^2$. Before diving into further calculations, it is convenient to write the elementary current \eqref{phi0HS} as follows.
\begin{equation}
    \mathcal{J}^{(0)}_k(y)^{B'}=(yq)\ J(k)\ g_k(y) \kbar^{B'} \, ,
\end{equation}
where $J(k)=\ub{qk}^{-1}$ and $g_k(y)=\ub{qk}(\ub{qk}+yq)^{-1}$. This way of writing the expression is justified to make contact with the notations defined in Appendix \ref{appBGC}, where a detailed analysis of the spin-1 case is presented. Moreover, it will be crucial to try to write objects from HS-SDYM in terms of their spin-1 counterpart. Actually, the only technical difference between HS-SDYM and SDYM lies in the fact that the action of the propagator \eqref{genpropag} seems rather intricate in the higher-spin case.\footnote{Indeed, the number operator acts on a function which is not a simple monomial.} For this reason, it is worth recalling the well-known  expressions of Berends--Giele currents in SDYM, and writing them in the language of generating functions, in order to compare them to their higher-spin equivalents.\footnote{The calculations are detailed in Appendix \ref{appBGC}.} Setting the variable $y$ to zero in the denominator of \eqref{phi0HS} is tantamount to projecting the generating function onto its spin-1 component. Therefore,
\begin{equation}
    J_k^{(0)}(y)^{B'}=\frac{yq}{\ub{qk}}\kbar^{B'} 
\end{equation}
is our spin-1 elementary current. The first- and second-order (or equivalently, the two- and three-point) currents are given, respectively, by
\begin{equation}
     J^{(1)}_{12}(y)^{B'}= (yq)\ q_B(\propagator{12})^{BB'} J(1,2)\ p_{12}^2 \, . \label{SDYMCo1}
\end{equation}
and
\begin{equation}
     J^{(2)}_{123}(y)^{B'}= (yq)\ q_B(\propagator{123})^{BB'} J(1,2,3)\ p_{123}^2 \, .
\end{equation}
Some shorthand notations are used. First, $p_{12\ldots n}=p_1+p_2+\ldots+p_n$, and the factor $J(1,2,\ldots,n)$ is defined as
\begin{equation}
    J(1,2,\ldots,n)=\frac{1}{\ub{q1}\ub{12}\ub{23}\ldots\ub{(n-1)n}\ub{nq}} \, .     \label{scalarJ} 
\end{equation}
The general pattern is quite straightforward, and the current for an arbitrary number of on-shell legs is given by
\begin{equation}
    J^{(n-1)}_{12\ldots n}(y)^{B'}=(yq)\ q_B(\propagator{12\ldots n})^{BB'}J(1,2,\ldots,n)\ p_{12\ldots n}^2 \, .\label{genphiYMgenfn}
\end{equation}
A crucial consequence of this formula is that all amplitudes, except the cubic one, vanish in SDYM because the factor $p_{12\ldots n}^2$ goes to zero when the on-shell limit is taken. Nevertheless, let us give the basic recipe to obtain an amplitude of order $n$ from the previous formula. It is done by removing the propagator $(\propagator{12\ldots n})^{BB'}$, contracting the reference spinor $q_B$ with one momentum factor of the polarization spin-tensor \eqref{negpol}, see equation \eqref{amplJeps}, and replacing the variable $y$ by the other momentum factor. As a consistency check, let us apply this procedure to \eqref{SDYMCo1}. Denoting the result of the previous recipe by $\ampl_3$, we obtain
\begin{equation}
   \ampl_3= \frac{\ub{q3}}{\ub{q1}} \frac{\ub{q3}}{\ub{q2}}\frac{\ub{12}\pb{12}}{\ub{12}}\, .
\end{equation}
The on-shell condition is $p_{12}^2=\ub{12}\pb{12}=0$. As for complex momenta, $\ub{12}$ and $\pb{12}$ are not related by complex conjugation, $\ub{12}$ can be set to zero while keeping $\ampl_3$ non-vanishing. Then, momentum conservation \eqref{momconsqk} ensures the following well-known result:
\begin{equation}
   \ampl_3= \frac{\pb{12}^3}{\pb{23}\pb{31}}\, .
\end{equation}
This expression is a particular case of \eqref{genericV} for $\Lambda>0$, where $\lbd_1=\lbd_2=+1$ and $\lbd_3=-1$. After this little digression on amplitudes, let us come back to calculating currents. The first-order current, computed from
\begin{equation}
    \mathcal{J}_{12}^{(1)}(y)^{B'}=\frac{1}{N+1}\pl_B(\propagator{12})^{BB'} \mathcal{J}^{(0)}_1(y)_{C'}\mathcal{J}^{(0)}_2(y)^{C'} \, ,
\end{equation}
takes the following form
\begin{equation}
    \mathcal{J}_{12}^{(1)}(y)^{B'}=(\propagator{12})^{BB'}J(1,2)\ p_{12}^2\ \left[\frac{1}{N+1}\pl_B\left[(yq)^2 g_1(y) g_2(y) \right]\right] \label{phi1tosolve} \, .
\end{equation}
The next step is to understand the action of the operator $(N+1)^{-1}$. As the quantities on which it acts are not simple monomials in $y^A$, it is better to use its integral form. Indeed, as shown in Appendix \ref{appintnumb}, the action of $(N+1)^{-1}$ on some generating function $F(x)$\footnote{A scalar variable $x\equiv yq$ has been defined for simplicity, but this does affect the result.} is 
\begin{equation}
    \frac{1}{N+1}\ F(x)=\int_0^1dt\ F(tx) \, ,
\end{equation}
This way of writing the operator can be used to derive the identity 
\begin{equation}
   \frac{1}{N+1}\ \pl_x(x^2 F(x))=xF(x) \label{idnumbint} \, ,
\end{equation}
which simplifies \eqref{phi1tosolve} to 
\begin{equation}
    \mathcal{J}_{12}^{(1)}(y)^{B'}=(yq)\ q_B(\propagator{12})^{BB'}J(1,2)\ p_{12}^2\  g_1(y) g_2(y)  \, . \label{2ptHS}
\end{equation}
This expression is exactly the one obtained in the SDYM case multiplied by some functions encoding the presence of spins greater than one. As another sample calculation, let us prove that the three-point higher-spin current has a similar structure. Using the algorithm presented in Appendix \ref{appBGC}, we find
$$
\begin{aligned}
    \mathcal{J}^{(2)}_{123}(y)^{B'}&=\frac{1}{N+1}\pl_B(\propagator{123})^{BB'}\left[\mathcal{J}^{(0)}_1(y)_{C'}\mathcal{J}^{(1)}_{23}(y)^{C'}+\mathcal{J}^{(1)}_{12}(y)_{C'}\mathcal{J}^{(0)}_3(y)^{C'}\right] \\
    &=\frac{1}{N+1}\pl_B(\propagator{12})^{BB'}(yq)^2\prod_{i=1}^3g_i(y)\left[q_C(p_{23})^{CC'}\pbar_{1C'}\ J(1)J(2,3)
    +q_C(p_{12})^{C}{}_{C'}\pbar_3^{C'}\ J(1,2)J(3)\right] \\
    &=(yq)\ \left(\prod_{i=1}^3g_i(y)\right) q_B(\propagator{123})^{BB'}J(1,2,3)\left[\frac{\ub{12}}{\ub{q2}} q_C(p_{23})^{CC'}\pbar_{1C'} 
    +\frac{\ub{23}}{\ub{q2}}q_C(p_{12})^{C}{}_{C'}\pbar_3^{C'}\right] \, .
\end{aligned}
$$
The product of $g_i(y)$ can be factored out and the result is just the SDYM current multiplied by these functions;\footnote{Appendix \ref{appBGC} contains a proof of the general Berends--Giele current in SDYM. The interested reader can refer to that part for the simplification of the terms in brackets.} 
\begin{equation}
    \mathcal{J}^{(2)}_{123}(y)^{B'}=(yq)\ \left(\prod_{i=1}^3\frac{\ub{qi}}{\ub{qi}+yq}\right) q_B(\propagator{123})^{BB'}J(1,2,3)\ p_{123}^2 \, .
    \label{phi2HSgengn}
\end{equation}
A consequence of this result is that the quartic amplitude vanishes in HS-SDYM. The fact that \eqref{idnumbint} can be applied to any function $F(x)$ is the key to conclude that all higher-point amplitudes vanish. Only the cubic ones survive (as long as momenta are complex). Indeed, given the expression for the theory restricted to spin 1 \eqref{genphiYMgenfn}, it is now straightforward to prove the generic formula of $\mathcal{J}^{(n-1)}(y)^{B'}$, which is
\begin{equation}
    \mathcal{J}^{(n-1)}_{12\ldots n}(y)^{B'}=(yq)\ \left(\prod_{i=1}^ng_i(y)\right) q_B(\propagator{12\ldots n})^{BB'}J(1,2,\ldots,n)\ p_{12\ldots n}^2\, .
    \label{phiHSgenfn}
\end{equation}
It is not difficult to deduce the general expression. Assuming that the expression of all currents up to order $n-2$ (or equivalently, up to $(n-1)$-point currents) are known, one can derive the current of order $n-1$ (also called the $n$-point current)
$$
\begin{aligned}
    \mathcal{J}^{(n-1)}_{12\ldots n}(y)^{B'}&=\frac{1}{N+1}\pl_B(\propagator{12\ldots n})^{BB'}\left[\mathcal{J}^{(0)}_1(y)_{C'}\mathcal{J}^{(n-2)}_{23\ldots n}(y)^{C'}+\ldots+\mathcal{J}^{(n-2)}_{12\ldots (n-1)}(y)_{C'}\mathcal{J}^{(1)}_{n}(y)^{C'}\right] \\
    &=\frac{1}{N+1}\pl_B(\propagator{12\ldots n})^{BB'}(yq)^2\left[\sum_{k=1}^{n-1}\left(q_C(p_{1\ldots k})^{C}{}_{C'}\ \left(\prod_{i=1}^k g_i(y)\right) J(1,\ldots,k)\right.\right. \\
    &\ \ \ \ \ \ \ \ \ \ \ \ \ \ \ \ \ \ \ \ \ \ \ \ \ \ \ \ \ \ \ \ \ \ \ \ \ \ \ \ \ \ \left.\left.\times q_D(p_{(k+1)\ldots n})^{DC'}\ \left(\prod_{j=k+1}^n g_j(y)\right) J(k+1,\ldots,n)\right)\right] \\
    &=\frac{1}{N+1}\pl_B(\propagator{12\ldots n})^{BB'}(yq)^2\prod_{l=1}^n g_l(y)\left[\sum_{k=1}^{n-1}\left(q_C(p_{1\ldots k})^{C}{}_{C'}\ J(1,\ldots,k)\right.\right. \\
    &\ \ \ \ \  \ \ \ \ \ \ \ \ \ \ \ \ \ \ \ \ \ \ \ \ \ \ \ \ \ \ \ \ \ \ \ \  \ \ \ \ \ \ \ \ \ \ \ \ \ \ \ \ \left.\left.\times q_D(p_{(k+1)\ldots n})^{DC'}\  J(k+1,\ldots,n)\right)\right] \\
    &=(yq)\ \left(\prod_{l=1}^ng_l(y)\right) q_B(\propagator{12\ldots n})^{BB'}J(1,\ldots,n)\left[\sum_{k=1}^{n-1}\frac{\ub{k(k+1)}}{\ub{qk}\ub{q(k+1)}} \sum_{1\leqslant i<j \leqslant n} \ub{qi}\ub{qj}\pb{ij} \right] \, ,
\end{aligned}
$$
where the sum in brackets is just the square norm of the sum of all on-shell momenta, $p_{12\ldots n}^2$. As expected, this factor causes most amplitudes to vanish for $n>2$, but the special case $n=2$ allows for the construction of cubic amplitudes 
\begin{equation}
    \ampl_3\sim \sum_{s_1+s_2-s_3=1} [12]^{s_1+s_2+s_3} [23]^{-s_1+s_2-s_3} [31]^{s_1-s_2-s_3}\, .
\end{equation}
These amplitudes form a subset of the Chiral HiSGRA amplitudes. They have the general form \eqref{genericV} with coefficients $1/\Gamma(s_1+s_2-s_3)=1$.

\section{Discussion and Conclusions} \label{ccl}

The main result of this paper is to confirm that the covariant equations of motion of Chiral HiSGRA lead to the correct cubic amplitudes. The cubic amplitudes completely determine the theory at least in flat space. This proves that the equations of \cite{Skvortsov:2022syz,Sharapov:2022faa,Sharapov:2022wpz,Sharapov:2022awp,Sharapov:2022nps,Sharapov:2023erv} do give an FDA for Chiral HiSGRA. As a simple extension of the above results to $n$-point amplitudes we computed them in a somewhat simpler HS-SDYM theory. The main advantage of HS-SDYM is that it admits a covariant and gauge-invariant action, which is an extension of Chalmers--Siegel action. We used the Berends--Giele's currents idea to prove that all tree-level amplitudes vanish. This result is new since the vanishing of tree-level amplitudes in Chiral HiSGRA does not immediately imply the same for HS-SDYM, even though all these theories are self-dual \cite{Ponomarev:2017nrr,Serrani:2025owx,Serrani:2025oaw} and should share the usual ``self-dual'' properties such as vanishing of tree-level amplitudes. 

It would be important to extend our results to the full Chiral HiSGRA. As a step towards this goal we have demonstrated that higher order contact vertices vanish in the special regime where the polarizations are chosen to share the same reference spinor. Another possible application is to ``self-dual'' holography, see e.g. \cite{Skvortsov:2018uru,Sharapov:2022awp,Jain:2024bza,Aharony:2024nqs,Skvortsov:2026gtq}. In AdS/CFT correspondence the leading energy pole of an AdS/CFT correlator corresponds to just a flat space amplitude in the same theory \cite{Maldacena:2011nz}. Since the tree-level amplitudes vanish in HS-SDYM (and also in Chiral HiSGRA, even though the latter was proven in the light-cone gauge), the leading pole of the corresponding AdS/CFT correlator must be absent as well, see e.g. \cite{Lipstein:2023pih, Chowdhury:2024dcy,Skvortsov:2026gtq} for the example of Yang--Mills theory in the holographic context. Another obvious extension is to construct Berends--Giele currents in AdS. This is straightforward for HS-SDYM since the theory is conformally-invariant \cite{Ponomarev:2017nrr}. Therefore, our results would directly give Berends--Giele currents of SDYM and HS-SDYM in $AdS_4$, if it had not been for homogeneous terms in the propagators \cite{Skvortsov:2026gtq}. It would also be interesting to compute the amplitudes from the twistor actions developed in \cite{Tran:2021ukl,Herfray:2022prf,Tran:2022tft,Tran:2025uad,Mason:2025pbz}. One can also think of the great variety of self-dual theories found recently in \cite{Serrani:2025owx}.

\section*{Acknowledgements}
I would like to thank Evgeny Skvortsov for proposing this project and for many useful discussions. This project was partially supported by European Research Council (ERC) under the European Union’s Horizon 2020 research and innovation (grant agreement No 101002551). The work of the author was supported by UMONS stipend ``Bourse d'encouragement doctorale FRIA/\-FRESH''.

\appendix

\section{From vectors to spinors} \label{appspin}

The aim of this section is to present the various conventions used in the main text, especially about the spinor notation, before introducing how to manipulate differential forms in this language in the next section. All these aspects are proper to four-dimensional theories, where it is convenient to convert all space-time indices into pairs of spinor indices. The isomorphism relating the algebras $\so{1,3}$ and $\slalg{2,\complex}$ allows the decomposition of vectors in Minkowski space-time into a basis of $(2\times2)$ matrices. Consider the momentum matrix
\begin{equation}
    p_{AA'} = p_\mu (\sigma^\mu)_{AA'}=
    \begin{pmatrix}
     -p^0+p^3 & p^1-ip^2\\
     p^1+ip^2 & -p^0-p^3\\ 
    \end{pmatrix} 
    \label{pAA} \, ,
\end{equation}
where the four basis matrices are the identity and the three Pauli matrices $(\sigma^\mu)_{AA'} = (\id_2,\sigma^i)$. Momenta and matrices are contracted by means of the mostly-plus Minkowski metric. The resulting matrix $p_{AA'}$ is Hermitian. Other conventions may lead to an anti-Hermitian matrix. Denoting the space of unprimed spinors by $S_+$ and the space of primed ones by $S_-$, one deduces that $p_{AA'}\in S_+\otimes S_-$. In the main text, \textit{helicity spinors} are used instead of such matrices. Helicity spinors can be defined by splitting the momentum matrix into a product of a column vector by a row vector. The crucial point is that all particles in our theory are massless ($p^2=0$), which implies $\det(p_{AA'})=0$. Indeed,
\begin{equation*}
    \det(p_{AA'}) = (p^0)^2 - (p^1)^2 - (p^2)^2 - (p^3)^2 = -p_\mu p^\mu = 0 \, .
\end{equation*}
Moreover, a singular $(2\times 2)$ matrix can be decomposed into an outer product of a column vector and a row vector as follows:
\begin{equation*}
    p_{AA'} = p_A \pbar_{A'}\ ,
\end{equation*}
where the helicity-spinors\footnote{Sometimes referred to as left and right spinors.} are commuting objects, in contrast to the ones used to describe fermions, which are Grassmann variables (anticommuting variables). The dual set $(\sigmabar^\mu)^{A'A} = (\id_2,-\sigma^i)$ can be introduced to define
\begin{equation}
    p^{A'A} = p_\mu (\sigmabar^\mu)^{A'A} \, .
    \label{xsigmaup}
\end{equation}
By using the identity $\trace(\sigma_\mu\sigmabar_\nu)=-2\eta_{\mu\nu}$,\footnote{This identity applies to the mostly-plus Minkowski metric.} one can invert the previous relations,
\begin{equation}
    p^\mu = -\frac{1}{2}\ p_{AA'}(\sigmabar^\mu)^{A'A} = -\frac{1}{2}\ p^{A'A}(\sigma^\mu)_{AA'} \, ,
\end{equation}
and translate the usual Minkowski inner product into spinor language, 
\begin{equation}
    p_\mu q^\mu = -\frac{1}{2}\ p_{AA'}q^{A'A}
    \label{xmuxmuxAAxAA} \, .
\end{equation}
Another important point is that primed and unprimed indices belong to independent representations. Indeed, the algebra $\so{1,3}$, is isomorphic to $\slalg{2,\complex}$ that has two fundamental representations --- unprimed and primed spinors. Moreover, the metric, which is an invariant symbol of the Lorentz group, gives rise to two new invariant symbols $\eps^{AB}$ ($\eps_{AB}$) and $\eps^{A'B'}$ ($\eps_{A'B'}$) which are used to raise and lower spinor indices
\begin{equation}
    \eta_{ab} = \eta_{AA'BB'} = \eps_{AB}\eps_{A'B'} \, .
\end{equation}
Unlike the Minkowski metric, these symbols are antisymmetric. Here are the raising and lowering rules for the unprimed indices
\[
y_A = y^B \eps_{BA} \text{ and } y^A = \eps^{AB} y_B \, .
\]
The rule is identical for primed indices. In matrix form, the epsilon symbol is
\begin{equation}
\eps^{AB}=\eps^{A'B'}=
\begin{pmatrix}
0 & 1 \\
-1 & 0
\end{pmatrix} \, .
\end{equation}
According to the conventions used to raise and lower indices, $\eps^{AB}=\eps_{AB}$ and, of course, $\eps^{A'B'}=\eps_{A'B'}$. One can figure out that $\eps_A{}^B=-\eps^B{}_A$ is the Kronecker delta $\delta^A_B$. The epsilon symbol defines an antisymmetric inner product, $\ub{pq}=p^Aq_A = -p_Aq^A$. Similarly, $[pq]=p^{A'}q_{A'}$. 

The spinor notation is very handy for the following reason. As the space of (left or right) spinors is 2-dimensional, one can make the following decomposition of an arbitrary spin-tensor $T_{AB}$.
\begin{equation}
    T_{AB} = T_{(AB)} + \frac{1}{2} \eps_{AB}\ T_C{}^C \, .
    \label{symmantisymmeps}
\end{equation}
$T_C{}^C$ is the trace of $T_{AB}$. The spin-tensor has simply been decomposed into its symmetric ($T_{(AB)}$) and antisymmetric ($T_{[AB]}$) components. The key point is that any antisymmetric tensor is proportional to the epsilon symbol. Then, only the symmetric part has a non-trivial structure. This rule will be the key to all the identities involving differential forms that will be shown in the next section.

A very useful tool for manipulating two-dimensional spinors is the Schouten identity (sometimes called Fierz identity),
\begin{equation}
    \phi_{AB}{}^{B}+\phi^B{}_{AB}{}+\phi_{B}{}^{B}{}_A =0\, .
\end{equation}
This is another way of writing the statement $\phi_{[ABC]}=0$. Of course, antisymmetrizing more than two indices gives a trivial result. In this paper, the identity is mostly used for momentum spinors and is therefore written as follows.
\begin{equation}
\ab{ij}\ab{kl}+\ab{ik}\ab{lj}+\ab{il}\ab{jk} = 0 \, .
    \label{schouten}
\end{equation}
Another way to derive this result is to write one of the spinors as a linear combination of the others. Indeed, three two-dimensional spinors cannot all be linearly independent. For example,
\begin{equation*}
    j\rangle = \frac{\ab{jl}}{\ab{kl}} k\rangle + \frac{\ab{jk}}{\ab{lk}} l\rangle \, ,
\end{equation*}
where the coefficients can be easily determined by taking some appropriate spinor products. Contracting this relation with $\langle i$ yields (\ref{schouten}). The identity can also be written in terms of $\eps$.
\begin{equation}
    \eps^{AB}\eps^{CD}+\eps^{AC}\eps^{DB}+\eps^{AD}\eps^{BC} = 0  \, .
    \label{schouteneps}
\end{equation}
Of course, the identity also holds for the other bracket product ($\rb{ij}\rb{kl}+\rb{ik}\rb{lj}+\rb{il}\rb{jk} = 0$).

\section{Differential forms} \label{appdiff}

The previous section, as well as the calculations in the main text, only involve the flat Minkowski background. Nevertheless, it may be worth discussing what happens when considering a curved space-time. First, vectors lying on the tangent bundle of the curved manifold are mapped to vectors in a copy of Minkowski space via the vierbein, $e^a_\mu\  v^\mu=v^a$. Then, the isomorphism that relates these vectors to $(2\times2)$ matrices is provided by a contraction with $\sigma_a^{AA'}$. The resulting object, $e^{AA'}_\mu=e^a_\mu\ \sigma_a^{AA'}$ (called sometimes the \textit{soldering form}) maps four-vectors to spinor matrices. The basic objects used in the main text are actually one-forms constructed from the soldering form; $e^{AA'}=d x^\mu \ e^{AA'}_\mu$. Here is a first simple identity satisfied by these vierbein one-forms:
\begin{equation}
    e_{AA'}\wedge e_{BB'} = \frac{1}{2} (\eps_{AB} H_{A'B'} + \eps_{A'B'} H_{AB}) \, ,
    \label{e^e}
\end{equation}
where $H_{AB} = e_{AC'}\wedge e_B{}^{C'}$ and $H_{A'B'} = e_{CA'}\wedge e^C{}_{B'}$ are symmetric and are referred to as self-dual and anti-self-dual two-forms.\footnote{There is indeed an isomorphism between self-dual objects and symmetric rank-2 spin-tensors (see \cite{Delfino_2015}).} This decomposition is valid for any object $T_{AA'|BB'}$ which is antisymmetric under the exchange $(AA')\leftrightarrow(BB')$. The proof results from applying (\ref{symmantisymmeps}) to both types of indices. Of course, primed and unprimed spinor indices can neither be contracted nor (anti)symmetrized. The object $H_{AB} = e_{AC'}\wedge e_B{}^{C'}$ is symmetric. There are three such two-forms, as well as three two-forms of type $H_{A'B'}$, which is consistent with the fact there are six independent two-forms in four dimensions. There are also four independent three-forms and and a single independent four-form (top form/volume form). The basis three-forms are denoted by $\hat{h}_{AA'}$, as they contain the same number of components as $e_{AA'}$. They can be constructed by using a one-form and a self-dual (or anti-self-dual) two-form. The following proportionality relation is expected,
\begin{equation}
    e_{AA'}\wedge H_{BC} \propto \eps_{AB}\ \hat{h}_{CA'} + \eps_{AC}\ \hat{h}_{BA'} \, ,
\end{equation}
as the self-dual two-form is symmetric. A contraction with an epsilon leads to $\hat{h}_{AA'} = e_{CA'}\wedge H_A{}^C$.\footnote{The coefficient has been chosen to be 1, but this is just a matter of convention.} Of course, there is no reason why the three forms could not be constructed from anti-self-dual two-forms. Indeed,
\begin{align*}
    \hat{h}_{AA'} & = e_{CA'}\wedge H^C{}_A = e_{CA'}\wedge e^C{}_{B'}\wedge e_A{}^{B'} = H_{A'C'}\wedge e_A{}^{C'} = - e_{AC'}\wedge H^{C'}{}_{A'} \, .
\end{align*}
The useful identities are therefore
\begin{equation}
    e_{AA'}\wedge H_{BC} = \frac{1}{3} (\eps_{AB} \hat{h}_{CA'} + \eps_{AC} \hat{h}_{BA'}) \, ,
\end{equation}
and
\begin{equation}
    e_{AA'}\wedge H_{B'C'} = -\frac{1}{3} (\eps_{A'B'} \hat{h}_{AC'} + \eps_{A'C'} \hat{h}_{AB'}) \, .
\end{equation}
The overall factor is deduced from contracting the expressions with the spinor metric. The previous techniques can be applied to write $H_{AB}\wedge H_{CD}$ as an index-free four-form multiplied by some epsilon factors carrying the indices. Knowing about the various symmetries, one can obtain 
\begin{equation}
H_{AB}\wedge H_{CD} = \frac{1}{6} (\eps_{AC} \eps_{BD} + \eps_{AD} \eps_{BC}) \ H_{XY}\wedge H^{XY} \, ,
\label{H^H}
\end{equation}
and
\begin{equation}
H_{A'B'}\wedge H_{C'D'} = \frac{1}{6} (\eps_{A'C'} \eps_{B'D'} + \eps_{A'D'} \eps_{B'C'}) \ H_{X'Y'}\wedge H^{X'Y'} \, .
\label{H'^H'}
\end{equation}
The quantity $H\wedge H \equiv H_{XY}\wedge H^{XY}=- H_{X'Y'}\wedge H^{X'Y'}$ is defined as the volume form. It can also be shown that
\begin{equation}
H_{AB}\wedge {H}_{C'D'} = 0 \, ,
\label{H^H'}
\end{equation}
for the simple reason that there is no invariant symbol $\eps$ mixing primed and unprimed indices, as required by Lorentz invariance. A last identity can be derived by considering $e_{AA'}\wedge \hat{h}_{BB'}$, which must be proportional to the volume form.
\begin{equation}
e_{AA'}\wedge \hat{h}_{BB'} = \frac{1}{4} \ \eps_{AB}\eps_{A'B'} \ H\wedge H \, .
\end{equation}
It can be obtained by making use of (\ref{e^e}), replacing the antisymmetric indices by epsilon symbols and taking a trace to find the overall coefficient.

In the main text, symmetric indices are often written with the same label. This is a very handy notation to use with spinor indices. A symmetric spin-tensor $T_{(A_1A_2)}=\frac{1}{2} (T_{A_1A_2}+T_{A_2A_1})$ is written $T_{AA}$. Of course, this can be generalized to any number of indices. $T_{(A_1...A_n)}=\frac{1}{n!} (T_{A_1...A_n}+\text{ all permutations of } (A_1,...,A_n))$ is denoted by $T_{A(n)}$. The notation $A(n)$ means that $T$ contains $n$ symmetric indices. With such a notation, the main identities derived in this section become
\begin{align}
    e_{AA'}\wedge H_{BB} & = \frac{2}{3}\  \epsilon_{AB}\ \hat{h}_{BA'} \, , \\
    e_{AA'}\wedge H_{B'C'} & = -\frac{2}{3}\  \epsilon_{A'B'}\ \hat{h}_{AB'} \ ,\\
H_{AA}\wedge H_{BB} & = \frac{1}{3}\  \eps_{AB} \eps_{AB} \ H \wedge H \, ,
\label{H^H} \\
H_{A'A'}\wedge H_{B'B'} & = -\frac{1}{3}\  \eps_{A'B'} \eps_{A'B'}  \ H \wedge H  \, ,
\label{H'^H'} \\
H_{AA}\wedge H_{B'B'} & = 0 \, ,
\label{H^H'} \\
e_{AA'}\wedge \hat{h}_{BB'} & = \frac{1}{4} \ \eps_{AB}\eps_{A'B'} \ H\wedge H \, .
\end{align}

\section{Amplitude constructed from $\uoc(h,C,C)$} \label{appuoc}

Let us write the total $\sigma_1$-expansion of the integrand in $\ampl_3$, keeping only the even powers since we are only interested in integer spins, 
\begin{equation}
     \left.\left[\frac{3\alpha}{[23]+\sigma_1\alpha}-\sigma_1 \left(\frac{\alpha}{[23]+\sigma_1\alpha}\right)^2\right]\right|_\text{even} = \sum_{s_1=0}^\infty \sigma_1^{2s_1-2} \left(\frac{\alpha}{[23]}\right)^{2s_1-1} (2s_1+1)\, .
\end{equation}
The amplitude now becomes
\begin{equation}
    \ampl_3=\frac{1}{24}\ e^{(\sigma_{2} \sigma_{3})^{-1}[23]}  (\sigma_{2}^{-1}[12]-\sigma_{3}^{-1}[31]) \sum_{s_1=0}^\infty \sigma_1^{2s_1-2} \int_{\Delta_{2}} \left(\frac{\alpha}{[23]}\right)^{2s_1-1} (2s_1+1) \, .
\end{equation}

Let us expand $\alpha^{2s_1-1}$. By defining $C=\sigma_2[31]$ and $A=\sigma_3[12]$, one obtains
\begin{equation}
    \int_{\Delta_2} \alpha^{2s_1-1} = \sum_{n=0}^{2s_1-1} \frac{(2s_1-1)!}{n!(2s_1-1-n)!}\ A^{2s_1-1-n}C^n  \int_{\Delta_2}(1-u)^n u^{2s_1-1-n} \, .
    \label{intalpha}
\end{equation}

One can follow similar steps for $\ucoc$ and $\ucco$. Here are the resulting amplitudes for all orderings of $hCC$. 
\begin{align}
    \ampl_3(\uocc)&=\frac{1}{24}\ e^{(\sigma_{2} \sigma_{3})^{-1}[23]}  (\sigma_{2}^{-1}\pb{12}-\sigma_{3}^{-1}\pb{31}) \sum_{s_1=1}^\infty \sigma_1^{2s_1-2} \int_{\Delta_{2}} \left(\frac{\alpha(v)}{[23]}\right)^{2s_1-1} (2s_1+1)\, ,\\
    \ampl_3(\ucoc)&=\frac{1}{24}\ e^{(\sigma_{2} \sigma_{3})^{-1}[23]}  (\sigma_{2}^{-1}\pb{12}+\sigma_{3}^{-1}\pb{31}) \sum_{s_1=1}^\infty \sigma_1^{2s_1-2} \int_{\Delta_{2}} \left(\frac{\alpha(1-v)}{[23]}\right)^{2s_1-1} (2s_1+1)\\ & \ \ \ +(u\leftrightarrow v)\, , \\
    \ampl_3(\ucco)&=-\frac{1}{24}\ e^{(\sigma_{2} \sigma_{3})^{-1}[23]}  (\sigma_{2}^{-1}\pb{12}-\sigma_{3}^{-1}\pb{31}) \sum_{s_1=1}^\infty \sigma_1^{2s_1-2} \int_{\Delta_{2}} \left(\frac{\alpha(u)}{[23]}\right)^{2s_1-1} (2s_1+1)\, .
\end{align}
Bearing in mind that $\alpha(x)=\sigma_3(1-x)[12]+\sigma_2x[31]$, the next step is to integrate expressions of the form $(1-x)^n x^{2s_1-1-n}$ over the simplex $\Delta_2$ defined by $\{0\le u\le v\le 1\}$, where $x$ can be either $u$ or $v$. Using
\begin{equation}
    \left\{\begin{aligned}
        \int_{\Delta_2}(1-u)^n u^{2s_1-1-n}& =\frac{(n+1)!(2s_1-1-n)!}{(2s_1+1)!}\, , \\
       \int_{\Delta_2}(1-v)^n v^{2s_1-1-n} & = \frac{n!(2s_1-n)!}{(2s_1+1)!}\, ,
    \end{aligned}
    \right.
\end{equation}


we obtain the following results for $\alpha^{2s_1-1}$. 
$$
\begin{aligned}
    &\int_{\Delta_2}[\alpha(u)]^{2s_1-1}=\sum_{n=0}^{2s_1-1} \frac{(n+1)}{2s}\ A^{2s_1-1-n}C^n \, , \\
    &\int_{\Delta_2}[\alpha(1-u)]^{2s_1-1}=\sum_{n=0}^{2s_1-1} \frac{(n+1)}{2s}\ A^n C^{2s_1-1-n} \, , \\
    &\int_{\Delta_2}[\alpha(v)]^{2s_1-1}=\sum_{n=0}^{2s_1-1} \frac{(2s-n)}{2s}\ A^{2s_1-1-n}C^n \, , \\
    &\int_{\Delta_2}[\alpha(1-v)]^{2s_1-1}=\sum_{n=0}^{2s_1-1} \frac{(2s-n)}{2s}\ A^n C^{2s_1-1-n} \, .
\end{aligned}
$$
It is easy to obtain $\alpha(1-u)$ from $\alpha(u)$. Indeed, from \eqref{intalpha}, one can see that interchanging the powers of $u$ and $(1-u)$ is equivalent to interchanging the powers of $A$ and $C$. Equation \eqref{totalamplitudeU} can be simplified by using the following identities.
\begin{align}
    &\int_{\Delta_2}[\alpha(u)]^{2s_1-1}=\int_{\Delta_2}[\alpha(1-v)]^{2s_1-1}\, , \\
    &\int_{\Delta_2}[\alpha(1-u)]^{2s_1-1}=\int_{\Delta_2}[\alpha(v)]^{2s_1-1}\, . \label{symmUV}
\end{align}
The amplitude \eqref{totalamplitudeU} now becomes 
\begin{equation}
    \ampl = \frac{1}{12} \ e^{(\sigma_{2}\sigma_{3})^{-1}[23]} \int_{\Delta_{2}} \left( \sigma_{2}^{-1}[12] F(v)+\sigma_{3}^{-1}[31]F(u)\right) \, .
\end{equation}
whose expansion yields the expression \eqref{ampltriliUexpanded}.

\section{Integral form of number operators}\label{appintnumb}




The action of the propagator in the higher-spin case \eqref{genpropag} might look rather intricate due to the expression of the higher-spin elementary currents being less trivial than in the spin-1 case.\footnote{Indeed, the number operator acts on a function which is not a simple monomial.} However, the problem can be solved by writing $(N+1)^{-1}$ as an integral acting on some generating function $F(x)$.\footnote{Let us temporarily rename $(yq)$ as $x$ to keep simple notation.} 
\begin{equation}
    \frac{1}{N+1}\ F(x)=\int_0^1dt\ F(tx) \, .
\end{equation}
The identity is proven by Taylor-expanding $F(x)$. Indeed, 
$$
\begin{aligned}
    \frac{1}{N+1}\ F(x)&=\frac{1}{N+1}\sum_{n=0}^\infty f_n x^n =\sum_{n=0}^\infty \frac{1}{n+1}f_n x^n =\sum_{n=0}^\infty \left(\int_0^1dt\ t^n\right)f_n x^n \\
    &=\int_0^1dt\ \sum_{n=0}^\infty f_n (tx)^n =\int_0^1dt\ F(tx)\, .
\end{aligned}
$$
In our case, the function $F(x)$ is $g_1(x) g_2(x)\ldots g_n(x)$, but it is better to keep it arbitrary. With the machinery that has been defined, the actions of both the derivative and the Euler operator can be evaluated together. 
\begin{align}
    \frac{1}{N+1}\ \pl_x(x^2 F(x))&=\int_0^1dt\  \frac{\pl}{\pl(tx)}((tx)^2 F(tx))\nonumber =\int_0^1dt\  \frac{1}{x}\frac{d}{dt}((tx)^2 F(tx))\nonumber \\
    &=\frac{1}{x}\Big[(tx)^2 F(tx)\Big]^1_0
    =xF(x) \label{trickNumberIntegral}\, .
\end{align}
This result is valid for any (differentiable) function $F(x)$. It follows that our two-point current takes a very simple structure.
\begin{equation}
    \mathcal{J}_{12}^{(1)}(y)^{B'}=(yq)\ q_B(\propagator{12})^{BB'}J(1,2)\ p_{12}^2\  g_1(y) g_2(y)  \, .
\end{equation}

\section{Berends--Giele recursion: a short review and application to SDYM} \label{appBGC}

The goal is to write the non-linear equations of motion for the gauge field in momentum space and solve them for each order in the coupling constant. This is equivalent to constructing Berends--Giele currents, which are sort of ``off-shell'' amplitudes. Such a derivation has been explicitly performed in \cite{Monteiro_2011} in the light-cone gauge. \cite{Krasnov:2016emc} mentions the general form of Berends--Giele currents in SDYM, which will be proven in the next pages. The SDYM action is 
\begin{equation}
    S=\trace\int \Psi^{AA} H_{AA}\wedge (dA-A\wedge A) \, .
\end{equation}
The gauge field $A$ --- a one-form taking values in a Lie algebra --- has been rescaled to hide the coupling constant $g$, so that the gauge field is $dA-A\wedge A$ instead of $dA-gA\wedge A$. The equation of motion for the field strength is 
\begin{equation}
    \pl_{AA'}A_{A}{}^{A'}=\left[A_{AA'},A_{A}{}^{A'}\right]\, .
\end{equation}
Defining 
\begin{equation}
    A=\sum_{n=0}^\infty A^{(n)} \, ,
\end{equation}
one obtains an expansion of the field in powers of the coupling constant. Hence, $A^{(n)}$ is of order $g^{n}$.\footnote{To be consistent with the action, one should say that $A^{(n)}$ is of order $g^{n+1}$, but we will assume that the right-hand side contains a coupling constant set to 1 and that $A^{(n)}$ is of order $g^{n}$.} As a first example, one can derive the first-order solution (in $g$) which corresponds to the 2-point Berends--Giele current. In momentum space, the equation to solve is
\begin{equation}
    k_{AA'}A^a_{A}{}^{A'}(k)=f^{abc}A^b_{AA'}(p_1)A^c_{A}{}^{A'}(p_2) \, ,
\end{equation}
where $f^{abc}$ are the structure constant of the considered Lie algebra. Contracting both sides of the equation with a momentum factor leads to\footnote{One can expect a term such as $k^B{}_{B'}k_{AA'}A^a_{B}{}^{A'}(k)$, but it can be shown that this term does not contribute by using the Fierz/Schouten identity and the Lorenz gauge condition, $k^{AA'}A_{AA'}(k)=0$.}
\begin{equation}
    k^B{}_{B'}k_{BA'}A^a_{A}{}^{A'}(k)=f^{abc}\ k^B{}_{B'}A^b_{BA'}(p_1)A^c_{A}{}^{A'}(p_2)\, .
\end{equation}
Then, one uses the identity $k_{BA'}k^B{}_{B'}=\eps_{A'B'}k^2$ to obtain
\begin{equation}
    A^{(1)a}_{AA'}(k) = f^{abc}\ \frac{k^B{}_{A'}}{k^2}A^{(0)b}_{BB'}(p_1)A^{(0)c}_{A}{}^{B'}(p_2) \, .
\end{equation}
The momentum $k$ is not on-shell, however it is equal to the sum of the on-shell momenta that appear on the right-hand side. Let us denote it by $p_{12}\equiv p_1+p_2$. A better expression of the solution is 
\begin{equation}
    A^{(1)a}_{AA'}(p_{12}) = f^{abc}\ (\propagator{12})^B{}_{A'}\ A^{(0)b}_{BB'}(p_1)A^{(0)c}_{A}{}^{B'}(p_2) \, . \label{A1YMindices}
\end{equation}
It is sufficient to consider only “color-stripped" currents. Hence, the color labels can be ignored from now on. 

Before going further in the calculation, let us give the general pattern that is followed by currents of generic order. A few sample calculations allows to guess that $A^{(n-1)}_{AA'}(p_{12\ldots n})$, which is the $(n-1)$-th order current carrying $n$ on-shell legs is obtained via the following algorithm:
\begin{equation}
    A^{(n-1)}_{AA'}(p_{12\ldots n}) =  (\propagator{12\ldots n})^B{}_{A'}\sum_{k=1}^{n-1} A^{(k-1)}_{BB'}(p_{12\ldots k})A^{(n-k-1)}_{A}{}^{B'}(p_{(k+1)\ldots n}) \, .\label{algoYMindices}
\end{equation}
Sample calculations will be performed in the next subsection, and the previous formula will be demonstrated by induction. One can construct by hand many currents by using the most elementary ones which have the form 
\begin{equation}
    A^{(0)}_{AA'}(k)=\frac{q_A\kbar_{A'}}{\ub{qk}}=q_A\kbar_{A'}J(k) \, ,
\end{equation}
where the factor $J(k)$ is just $\frac{1}{\ub{qk}}$. The general result is  
\begin{equation}
    A^{(n-1)}_{AA'}(p_{12\ldots n}) = q_Aq_B(\propagator{12\ldots n})^B{}_{A'}\ J(1,2,\ldots,n)\ p_{12\ldots n}^2 \, ,
    \label{genBGYMindices}
\end{equation}
where $p_{12\ldots n}=p_1+p_2+\ldots+p_n$ and the factor $J$ is defined by
\begin{equation}
    J(1,2,\ldots,n)=\frac{1}{\ub{q1}\ub{12}\ub{23}\ldots\ub{(n-1)n}\ub{nq}} \, .     \label{scalarJ} 
\end{equation}
The purpose of the relation \eqref{genBGYMindices} is to construct an amplitude with $n+1$ on-shell legs. In the case of SDYM, it is known that all amplitudes with more than 3 legs vanish. Indeed, momentum conservation imposes
$p_{12\ldots n}^2=0$ that leads to a vanishing amplitude, except for $n=2$, as shown in the following calculation. Let us rewrite the first-order current as 
$$
\begin{aligned}
    A^{(1)}_{AA'}(p_{12}) & = q_Aq_B(\propagator{12})^B{}_{A'}\ J(1,2)\ p_{12}^2 \\
    & = q_Aq_B(\propagator{12})^B{}_{A'} \frac{1}{\ub{q1}\ub{12}\ub{2q}} \ub{12}[12] \\
    & = q_Aq_B(\propagator{12})^B{}_{A'} \frac{1}{\ub{q1}\ub{q2}} [21] \, .
\end{aligned}
$$
As $p_{12}^2$ contains only one term, the factor $\ub{12}$ does not contribute and can be set to zero.\footnote{This does not affect the amplitude provided the momenta are complex; if momenta were real, $\ub{kl}$ and $[kl]$ would be related by complex conjugation.} Therefore, on-shell, we have $p_{12}^2=0$ without affecting the amplitude, which is computed by (i) removing the propagator and (ii) contracting the result with a negative-helicity spin-tensor, namely $p_3^Ap_3^A$. The resulting cubic amplitude is
\begin{equation}
    \ampl_3=\frac{\ub{q3}^2}{\ub{q1}\ub{q2}}[21]=\frac{[12]^3}{[23][13]} \, .
\end{equation}

Instead of using many indices, it is more elegant to prove \eqref{genBGYMindices} by using generating functions, even for spin 1. In the next section, such machinery will be defined to compute spin-1 currents and prove the generic formula. This proof consists of a major building block for the higher-spin generalization. 

\subsection{Derivation of some spin-1 currents using generating functions}

It is rather convenient, even for spin-1 fields, to use generating functions, as they are most practical when dealing with higher spins, like in the main text. To make contact with notations from the main text, the currents will be denoted by $J^{(0)}(y)^{B'}$. Here, the functions contain only one $y$. Indeed, restricting \eqref{phi0HS} to spin 1 is done by setting $y$ to zero in the denominator. The elementary current is thus
\begin{equation}
    J^{(0)}(y)^{B'}=\frac{yq}{\ub{qk}}\kbar^{B'} \ .
\end{equation}
The number of $y$'s involved in each calculation is definite, and the action of the operator $(N+1)^{-1}$ is trivial. As a warm-up, let us compute the first-order (or, equivalently, the two-point) current. Following the recipe prescribed by \eqref{A1YMindices}, 
$$
\begin{aligned}
    J^{(1)}_{12}(y)^{B'}&=\frac{1}{N+1}\pl_B(\propagator{12})^{BB'}J^{(0)}_1(y)_{C'}J^{(0)}_2(y)^{C'} =\frac{1}{N+1}\pl_B(\propagator{12})^{BB'}\frac{yq}{\ub{q1}}\frac{yq}{\ub{q2}}[21]\\
    &=\frac{2}{N+1}(yq)\ q_B(\propagator{12})^{BB'}\frac{[21]}{\ub{q1}\ub{q2}}= (yq)\ q_B(\propagator{12})^{BB'} J(1,2)\ p_{12}^2 \ .
\end{aligned}
$$
The subscript is a shorthand notation to indicate which on-shell momenta the current depends on. To compute the next current, one needs two terms:\footnote{This reflects the contribution of the two channels needed to compute a four-point color-ordered amplitude in Yang--Mills theory. Indeed, after some rearrangement of the terms in the gauge-invariant amplitude, it is possible to show that only the trivalent planar diagrams (namely, the s-channel and the t-channel) are required.} 
$$
\begin{aligned}
    J^{(2)}_{123}(y)^{B'}&=\frac{1}{N+1}\pl_B(\propagator{123})^{BB'}\left[J^{(0)}_1(y)_{C'}J^{(1)}_{23}(y)^{C'}+J^{(1)}_{12}(y)_{C'}J^{(0)}_3(y)^{C'}\right] \\
    &=\frac{1}{N+1}\pl_B(\propagator{123})^{BB'}(yq)^2\left[q_C(p_{23})^{CC'}\pbar_{1C'}\ J(1)J(2,3)
    +q_C(p_{12})^{C}{}_{C'}\pbar_3^{C'}\ J(1,2)J(3)\right] \\
    &=(yq)\ q_B(\propagator{123})^{BB'}\left[ q_C(p_{23})^{CC'}\pbar_{1C'}\ J(1)J(2,3)
    +q_C(p_{12})^{C}{}_{C'}\pbar_3^{C'}\ J(1,2)J(3)\right]\\
    &=(yq)\ q_B(\propagator{123})^{BB'}J(1,2,3)\left[\frac{\ub{12}}{\ub{q2}} q_C(p_{23})^{CC'}\pbar_{1C'} 
    +\frac{\ub{23}}{\ub{q2}}q_C(p_{12})^{C}{}_{C'}\pbar_3^{C'}\right] \, .
\end{aligned}
$$
The last line is obtained by relating $J(1)J(2,3)$ and $J(1,2)J(3)$ to $J(1,2,3)$. It is straightforward to write the generic formula for $1<k<n-1$,
\begin{equation}
    J(1,\ldots,k)J(k+1,\ldots,n)=J(1,\ldots,n)\ \frac{\ub{k(k+1)}}{\ub{kq}\ub{q(k+1)}} \, . 
\end{equation}
The last step is to show that the terms in brackets are equal to $p_{123}^2$. 
$$
\begin{aligned}
    [\ldots]&=\frac{1}{\ub{q2}}\left[\ub{12} (\ub{q2}\pb{12}+\ub{q3}\pb{13})
    +\ub{23}(\ub{q1}\pb{13}+\ub{q2}\pb{23})\right] \\
    &=\ub{12}\pb{12}+\ub{23}\pb{23}+\frac{\pb{13}}{\ub{q2}} (\ub{q3}\ub{12}
    +\ub{q1}\ub{23}) \\
    &=\ub{12}\pb{12}+\ub{23}\pb{23}+\ub{13}\pb{13}\\
    &=p_{123}^2
\end{aligned}
$$
The final result has the expected structure;
\begin{equation}
    J^{(2)}_{123}(y)^{B'}=(yq)\ q_B(\propagator{123})^{BB'}J(1,2,3)\ p_{123}^2 \, .\label{phi2YMgenfn}
\end{equation}
The higher the order of the current, the more cumbersome the previous manipulations become. The next section presents the general proof for the expression of Berends--Giele currents in SDYM.


\section{All-order Construction of Berends--Giele currents in SDYM}

Let us give an inductive proof of the general expression \eqref{genphiYMgenfn}, starting from the expression
\begin{equation}
    J^{(n-1)}_{12\ldots n}(y)^{B'}= \frac{1}{N+1} \pl_B(\propagator{12\ldots n})^{BB'}\ \sum_{k=1}^{n-1} J^{(k-1)}_{12\ldots k}(y)_{C'}J^{(n-k-1)}_{(k+1)\ldots n}(y)^{C'} \label{algoHSYMgenfn} \, ,
\end{equation}
and assuming that for $k<n$, the currents follow the expected pattern
\begin{equation}
    J^{(k-1)}_{12\ldots k}(y)^{B'}=(yq)\ q_B(\propagator{12\ldots k})^{BB'}J(1,2,\ldots,k)\ p_{12\ldots k}^2 \, ,\label{genphiYMgenfnAPP} 
\end{equation}
where 
\begin{align}
    &J(1,2,\ldots,k)=\frac{1}{\ub{q1}\ub{12}\ub{23}\ldots\ub{(k-1)k}\ub{kq}} \, ,\\
    &p_{12\ldots k}=p_1+p_2+\ldots+p_k\, ,\\
    &\propagator{12\ldots k}=p_{12\ldots k}/p_{12\ldots k}^2 \, .
\end{align}
First, using the simple relation
\begin{equation*}
    J(1,\ldots,k)J(k+1,\ldots,n)=J(1,\ldots,n)\ \frac{\ub{k(k+1)}}{\ub{kq}\ub{q(k+1)}} 
\end{equation*}
allows to write the n-point current as
\begin{equation*}
    \phi^{(n-1)}_{12\ldots n}(y)^{B'}= (yq)\ q_B(\propagator{12\ldots n})^{BB'}\ J(1,\ldots,n) \sum_{k=1}^{n-1} \frac{\ub{k(k+1)}}{\ub{qk}\ub{q(k+1)}}\ \sum_{i=1}^{k}\sum_{j=k+1}^{n} \ub{qi}\ub{qj}\pb{ij} \, .
\end{equation*}
The main part of the proof consists of proving that the sum following $J(1,\ldots,n)$ is equal to $p_{12\ldots n}^2$, or more explicitly, 
\begin{equation}
    \sum_{1\leqslant i<j\leqslant n} \ub{ij}\pb{ij} \, .
\end{equation}
From the calculation of $J_{123}^{(2)}(y)^{B'}$, one can guess that the general procedure is to factor out each $[ij]$ and use the Schouten identity \eqref{schouten} (as many times as necessary) with the angle-bracket products, until forming $\ub{ij}$. This can be achieved by massaging summation indices, without using any property of the spinor brackets. As a result, the following equality holds
\begin{equation}
    \sum_{k=1}^{n-1} \frac{\ub{k(k+1)}}{\ub{qk}\ub{q(k+1)}}\ \sum_{i=1}^{k}\sum_{j=k+1}^{n} \ub{qi}\ub{qj}\pb{ij} = \sum_{1\leqslant i<j\leqslant n}\pb{ij}\left(\ub{qi}\ub{qj}\sum_{k=i}^{j-1} \frac{\ub{k(k+1)}}{\ub{qk}\ub{q(k+1)}}\right) \label{qiqjij} \, .
\end{equation}
The last step of the proof consists of establishing that
\begin{equation} 
    \ub{qi}\ub{qj}\sum_{k=i}^{j-1} \frac{\ub{k(k+1)}}{\ub{qk}\ub{q(k+1)}}=\ub{ij} \, .
    \label{induc}
\end{equation}
This can be demonstrated by induction. Of course, when $j=i+1$, the equality is trivially satisfied. Let us assume that the equality is satisfied for some integer $j=l>2$ and illustrate that it holds for $j=l+1$. Equation \eqref{induc} is thus our induction hypothesis. It is assumed to hold for some $j$ and the objective is to show its validity for $j+1$. 
$$
\begin{aligned}
    \sum_{k=i}^{j} \frac{\ub{k(k+1)}}{\ub{qk}\ub{q(k+1)}}&=\sum_{k=i}^{j-1} \frac{\ub{k(k+1)}}{\ub{qk}\ub{q(k+1)}}+\frac{\ub{j(j+1)}}{\ub{qj}\ub{q(j+1)}} \\
    &=\frac{\ub{ij}}{\ub{qi}\ub{qj}} +\frac{\ub{j(j+1)}}{\ub{qj}\ub{q(j+1)}} \\
    &=\frac{1}{\ub{qj}}\left(\frac{\ub{q(j+1)}\ub{ij}+\ub{qi}\ub{j(j+1)}}{\ub{qi}\ub{q(j+1)}}\right) \\
    &=\frac{1}{\ub{qj}}\left(\frac{\ub{qj}\ub{i(j+1)}}{\ub{qi}\ub{q(j+1)}}\right) \\
    &=\frac{\ub{i(j+1)}}{\ub{qi}\ub{q(j+1)}}\, .
\end{aligned}
$$
The second line follows from the induction hypothesis. The fourth one is the consequence of the Schouten identity. This completes the proof of Formula \eqref{genphiYMgenfn}.

\footnotesize
\providecommand{\href}[2]{#2}\begingroup\raggedright\endgroup


\begin{thebibliography}{10}

\bibitem{Bekaert:2022poo}
X.~Bekaert, N.~Boulanger, A.~Campoleoni, M.~Chiodaroli, D.~Francia,
  M.~Grigoriev, E.~Sezgin, and E.~Skvortsov, ``{Snowmass White Paper: Higher
  Spin Gravity and Higher Spin symmetry},''
  \href{http://arxiv.org/abs/2205.01567}{{\ttfamily arXiv:2205.01567
  [hep-th]}}.

\bibitem{Blencowe:1988gj}
M.~Blencowe, ``{A Consistent Interacting Massless Higher Spin Field Theory in
  $D$ = (2+1)},''
{\em Class.Quant.Grav.} {\bfseries 6} (1989) 443.

\bibitem{Bergshoeff:1989ns}
E.~Bergshoeff, M.~P. Blencowe, and K.~S. Stelle, ``{Area Preserving
  Diffeomorphisms and Higher Spin Algebra},''
{\em Commun. Math. Phys.} {\bfseries 128} (1990) 213.

\bibitem{Campoleoni:2010zq}
A.~Campoleoni, S.~Fredenhagen, S.~Pfenninger, and S.~Theisen, ``{Asymptotic
  symmetries of three-dimensional gravity coupled to higher-spin fields},''
  {\em JHEP} {\bfseries 1011} (2010) 007,
\href{http://arxiv.org/abs/1008.4744}{{\ttfamily arXiv:1008.4744 [hep-th]}}.

\bibitem{Henneaux:2010xg}
M.~Henneaux and S.-J. Rey, ``{Nonlinear $W_{\infty}$ as Asymptotic Symmetry of
  Three-Dimensional Higher Spin Anti-de Sitter Gravity},'' {\em JHEP}
  {\bfseries 1012} (2010) 007,
\href{http://arxiv.org/abs/1008.4579}{{\ttfamily arXiv:1008.4579 [hep-th]}}.

\bibitem{Grigoriev:2020lzu}
M.~Grigoriev, K.~Mkrtchyan, and E.~Skvortsov, ``{Matter-free higher spin
  gravities in 3D: Partially-massless fields and general structure},''
  \href{http://dx.doi.org/10.1103/PhysRevD.102.066003}{{\em Phys. Rev. D}
  {\bfseries 102} no.~6, (2020) 066003},
  \href{http://arxiv.org/abs/2005.05931}{{\ttfamily arXiv:2005.05931
  [hep-th]}}.

\bibitem{Pope:1989vj}
C.~N. Pope and P.~K. Townsend, ``{Conformal Higher Spin in (2+1)-dimensions},''
{\em Phys. Lett.} {\bfseries B225} (1989) 245--250.

\bibitem{Fradkin:1989xt}
E.~S. Fradkin and V.~{\relax Ya}. Linetsky, ``{A Superconformal Theory of
  Massless Higher Spin Fields in $D$ = (2+1)},'' {\em Mod. Phys. Lett.}
  {\bfseries A4} (1989) 731.
[Annals Phys.198,293(1990)].

\bibitem{Grigoriev:2019xmp}
M.~Grigoriev, I.~Lovrekovic, and E.~Skvortsov, ``{New Conformal Higher Spin
  Gravities in $3d$},'' {\em JHEP} {\bfseries 01} (2020) 059,
\href{http://arxiv.org/abs/1909.13305}{{\ttfamily arXiv:1909.13305 [hep-th]}}.

\bibitem{Alkalaev:2014qpa}
K.~B. Alkalaev, ``{Global and local properties of AdS$_{2}$ higher spin
  gravity},'' \href{http://dx.doi.org/10.1007/JHEP10(2014)122}{{\em JHEP}
  {\bfseries 10} (2014) 122}, \href{http://arxiv.org/abs/1404.5330}{{\ttfamily
  arXiv:1404.5330 [hep-th]}}.

\bibitem{Alkalaev:2019xuv}
K.~Alkalaev and X.~Bekaert, ``{Towards higher-spin AdS$_2$/CFT$_1$
  holography},'' \href{http://dx.doi.org/10.1007/JHEP04(2020)206}{{\em JHEP}
  {\bfseries 04} (2020) 206}, \href{http://arxiv.org/abs/1911.13212}{{\ttfamily
  arXiv:1911.13212 [hep-th]}}.

\bibitem{Alkalaev:2020kut}
K.~Alkalaev and X.~Bekaert, ``{On BF-type higher-spin actions in two
  dimensions},'' \href{http://dx.doi.org/10.1007/JHEP05(2020)158}{{\em JHEP}
  {\bfseries 05} (2020) 158}, \href{http://arxiv.org/abs/2002.02387}{{\ttfamily
  arXiv:2002.02387 [hep-th]}}.

\bibitem{Sharapov:2024euk}
A.~Sharapov, E.~Skvortsov, and A.~Sukhanov, ``{Matter-coupled higher spin
  gravities in 3d: no- and yes-go results},''
  \href{http://dx.doi.org/10.1007/JHEP04(2025)155}{{\em JHEP} {\bfseries 04}
  (2025) 155}, \href{http://arxiv.org/abs/2409.12830}{{\ttfamily
  arXiv:2409.12830 [hep-th]}}.

\bibitem{Bekaert:2025azj}
X.~Bekaert, A.~Sharapov, and E.~Skvortsov, ``{Higher-Spin Poisson Sigma Models
  and Holographic Duality for SYK Models},''
  \href{http://arxiv.org/abs/2509.19964}{{\ttfamily arXiv:2509.19964
  [hep-th]}}.

\bibitem{Segal:2002gd}
A.~Y. Segal, ``{Conformal higher spin theory},'' {\em Nucl. Phys.} {\bfseries
  B664} (2003) 59--130,
\href{http://arxiv.org/abs/hep-th/0207212}{{\ttfamily arXiv:hep-th/0207212
  [hep-th]}}.

\bibitem{Tseytlin:2002gz}
A.~A. Tseytlin, ``{On limits of superstring in $AdS_5\times S^5$},'' {\em
  Theor. Math. Phys.} {\bfseries 133} (2002) 1376--1389,
  \href{http://arxiv.org/abs/hep-th/0201112}{{\ttfamily arXiv:hep-th/0201112
  [hep-th]}}.
[Teor. Mat. Fiz.133,69(2002)].

\bibitem{Bekaert:2010ky}
X.~Bekaert, E.~Joung, and J.~Mourad, ``{Effective action in a higher-spin
  background},'' {\em JHEP} {\bfseries 02} (2011) 048,
\href{http://arxiv.org/abs/1012.2103}{{\ttfamily arXiv:1012.2103 [hep-th]}}.

\bibitem{Basile:2022nou}
T.~Basile, M.~Grigoriev, and E.~Skvortsov, ``{Covariant action for conformal
  higher spin gravity},''
  \href{http://dx.doi.org/10.1088/1751-8121/aceeca}{{\em J. Phys. A} {\bfseries
  56} no.~38, (2023) 385402}, \href{http://arxiv.org/abs/2212.10336}{{\ttfamily
  arXiv:2212.10336 [hep-th]}}.

\bibitem{Bengtsson:1983pg}
A.~K.~H. Bengtsson, I.~Bengtsson, and L.~Brink, ``{Cubic interaction terms for
  arbitrarily extended Supermultiplets},''
{\em Nucl. Phys.} {\bfseries B227} (1983) 41.

\bibitem{Bengtsson:1983pd}
A.~K.~H. Bengtsson, I.~Bengtsson, and L.~Brink, ``{Cubic interaction terms for
  arbitrary spin},''
{\em Nucl. Phys.} {\bfseries B227} (1983) 31.

\bibitem{Bengtsson:1986kh}
A.~K.~H. Bengtsson, I.~Bengtsson, and N.~Linden, ``{Interacting Higher Spin
  Gauge Fields on the Light Front},''
{\em Class. Quant. Grav.} {\bfseries 4} (1987) 1333.

\bibitem{Metsaev:1991nb}
R.~R. Metsaev, ``{$S$ matrix approach to massless higher spins theory. 2: The
  Case of internal symmetry},''
{\em Mod. Phys. Lett.} {\bfseries A6} (1991) 2411--2421.

\bibitem{Metsaev:1991mt}
R.~R. Metsaev, ``{Poincare invariant dynamics of massless higher spins: Fourth
  order analysis on mass shell},''
{\em Mod. Phys. Lett.} {\bfseries A6} (1991) 359--367.

\bibitem{Bengtsson:2014qza}
A.~K.~H. Bengtsson, ``{A Riccati type PDE for light-front higher helicity
  vertices},'' \href{http://dx.doi.org/10.1007/JHEP09(2014)105}{{\em JHEP}
  {\bfseries 09} (2014) 105},
\href{http://arxiv.org/abs/1403.7345}{{\ttfamily arXiv:1403.7345 [hep-th]}}.

\bibitem{Conde:2016izb}
E.~Conde, E.~Joung, and K.~Mkrtchyan, ``{Spinor-Helicity Three-Point Amplitudes
  from Local Cubic Interactions},''
  \href{http://dx.doi.org/10.1007/JHEP08(2016)040}{{\em JHEP} {\bfseries 08}
  (2016) 040},
\href{http://arxiv.org/abs/1605.07402}{{\ttfamily arXiv:1605.07402 [hep-th]}}.

\bibitem{Fronsdal:1978rb}
C.~Fronsdal, ``Massless fields with integer spin,''
{\em Phys. Rev.} {\bfseries D18} (1978) 3624.

\bibitem{Ponomarev:2016lrm}
D.~Ponomarev and E.~D. Skvortsov, ``{Light-Front Higher-Spin Theories in Flat
  Space},'' {\em J. Phys.} {\bfseries A50} no.~9, (2017) 095401,
\href{http://arxiv.org/abs/1609.04655}{{\ttfamily arXiv:1609.04655 [hep-th]}}.

\bibitem{Penrose:1965am}
R.~Penrose, ``{Zero rest mass fields including gravitation: Asymptotic
  behavior},''
\href{http://dx.doi.org/10.1098/rspa.1965.0058}{{\em Proc. Roy. Soc. Lond.}
  {\bfseries A284} (1965) 159}.

\bibitem{Hughston:1979tq}
L.~P. Hughston, R.~S. Ward, M.~G. Eastwood, M.~L. Ginsberg, A.~P. Hodges, S.~A.
  Huggett, T.~R. Hurd, R.~O. Jozsa, R.~Penrose, A.~Popovich, {\em et~al.},
  eds., {\em {Advances in twistor theory}}.
\newblock
1979.
\newblock

\bibitem{Eastwood:1981jy}
M.~G. Eastwood, R.~Penrose, and R.~O. Wells, ``{Cohomology and Massless
  Fields},''
\href{http://dx.doi.org/10.1007/BF01942327}{{\em Commun. Math. Phys.}
  {\bfseries 78} (1981) 305--351}.

\bibitem{Woodhouse:1985id}
N.~M.~J. Woodhouse, ``{Real methods in twistor theory},''
\href{http://dx.doi.org/10.1088/0264-9381/2/3/006}{{\em Class. Quant. Grav.}
  {\bfseries 2} (1985) 257--291}.

\bibitem{Sharapov:2022faa}
A.~Sharapov, E.~Skvortsov, A.~Sukhanov, and R.~Van~Dongen, ``{Minimal model of
  Chiral Higher Spin Gravity},''
  \href{http://arxiv.org/abs/2205.07794}{{\ttfamily arXiv:2205.07794
  [hep-th]}}.

\bibitem{Sharapov:2022wpz}
A.~Sharapov, E.~Skvortsov, and R.~Van~Dongen, ``{Chiral higher spin gravity and
  convex geometry},''
  \href{http://dx.doi.org/10.21468/SciPostPhys.14.6.162}{{\em SciPost Phys.}
  {\bfseries 14} no.~6, (2023) 162},
  \href{http://arxiv.org/abs/2209.01796}{{\ttfamily arXiv:2209.01796
  [hep-th]}}.

\bibitem{Sharapov:2022awp}
A.~Sharapov and E.~Skvortsov, ``{Chiral higher spin gravity in (A)dS4 and
  secrets of Chern{\textendash}Simons matter theories},''
  \href{http://dx.doi.org/10.1016/j.nuclphysb.2022.115982}{{\em Nucl. Phys. B}
  {\bfseries 985} (2022) 115982},
  \href{http://arxiv.org/abs/2205.15293}{{\ttfamily arXiv:2205.15293
  [hep-th]}}.

\bibitem{Sharapov:2022nps}
A.~Sharapov, E.~Skvortsov, A.~Sukhanov, and R.~Van~Dongen, ``{More on Chiral
  Higher Spin Gravity and convex geometry},''
  \href{http://dx.doi.org/10.1016/j.nuclphysb.2023.116152}{{\em Nucl. Phys. B}
  {\bfseries 990} (2023) 116152},
  \href{http://arxiv.org/abs/2209.15441}{{\ttfamily arXiv:2209.15441
  [hep-th]}}.

\bibitem{Sharapov:2023erv}
A.~Sharapov, E.~Skvortsov, and R.~Van~Dongen, ``{Strong homotopy algebras for
  chiral higher spin gravity via Stokes theorem},''
  \href{http://dx.doi.org/10.1007/JHEP06(2024)186}{{\em JHEP} {\bfseries 06}
  (2024) 186}, \href{http://arxiv.org/abs/2312.16573}{{\ttfamily
  arXiv:2312.16573 [hep-th]}}.

\bibitem{Ponomarev:2017nrr}
D.~Ponomarev, ``{Chiral Higher Spin Theories and Self-Duality},'' {\em JHEP}
  {\bfseries 12} (2017) 141,
\href{http://arxiv.org/abs/1710.00270}{{\ttfamily arXiv:1710.00270 [hep-th]}}.

\bibitem{Skvortsov:2018jea}
E.~D. Skvortsov, T.~Tran, and M.~Tsulaia, ``{Quantum Chiral Higher Spin
  Gravity},'' {\em Phys. Rev. Lett.} {\bfseries 121} no.~3, (2018) 031601,
\href{http://arxiv.org/abs/1805.00048}{{\ttfamily arXiv:1805.00048 [hep-th]}}.

\bibitem{Skvortsov:2020wtf}
E.~Skvortsov, T.~Tran, and M.~Tsulaia, ``{More on Quantum Chiral Higher Spin
  Gravity},'' {\em Phys. Rev.} {\bfseries D101} no.~10, (2020) 106001,
\href{http://arxiv.org/abs/2002.08487}{{\ttfamily arXiv:2002.08487 [hep-th]}}.

\bibitem{Skvortsov:2020gpn}
E.~Skvortsov and T.~Tran, ``{One-loop Finiteness of Chiral Higher Spin
  Gravity},''
\href{http://arxiv.org/abs/2004.10797}{{\ttfamily arXiv:2004.10797 [hep-th]}}.

\bibitem{Tsulaia:2022csz}
M.~Tsulaia and D.~Weissman, ``{Supersymmetric quantum chiral higher spin
  gravity},'' \href{http://dx.doi.org/10.1007/JHEP12(2022)002}{{\em JHEP}
  {\bfseries 12} (2022) 002}, \href{http://arxiv.org/abs/2209.13907}{{\ttfamily
  arXiv:2209.13907 [hep-th]}}.

\bibitem{Skvortsov:2022syz}
E.~Skvortsov and R.~Van~Dongen, ``{Minimal models of field theories: Chiral
  Higher Spin Gravity},'' \href{http://arxiv.org/abs/2204.10285}{{\ttfamily
  arXiv:2204.10285 [hep-th]}}.

\bibitem{Singh:1974qz}
L.~P.~S. Singh and C.~R. Hagen, ``Lagrangian formulation for arbitrary spin. 1.
  the boson case,''
{\em Phys. Rev.} {\bfseries D9} (1974) 898--909.

\bibitem{Krasnov:2021nsq}
K.~Krasnov, E.~Skvortsov, and T.~Tran, ``{Actions for Self-dual Higher Spin
  Gravities},''
\href{http://arxiv.org/abs/2105.12782}{{\ttfamily arXiv:2105.12782 [hep-th]}}.

\bibitem{Tran:2025yzd}
T.~Tran, ``{Self-dual pp-wave solutions in chiral higher-spin gravity},''
  \href{http://dx.doi.org/10.1007/JHEP03(2025)041}{{\em JHEP} {\bfseries 03}
  (2025) 041}, \href{http://arxiv.org/abs/2501.06445}{{\ttfamily
  arXiv:2501.06445 [hep-th]}}.

\bibitem{Skvortsov:2025ohi}
E.~Skvortsov and Y.~Yin, ``{Higher-spins on Taub-NUT and higher-spin
  Taub-NUT},'' \href{http://dx.doi.org/10.1007/JHEP12(2025)099}{{\em JHEP}
  {\bfseries 12} (2025) 099}, \href{http://arxiv.org/abs/2508.18804}{{\ttfamily
  arXiv:2508.18804 [hep-th]}}.

\bibitem{Skvortsov:2024rng}
E.~Skvortsov and Y.~Yin, ``{Low spin solutions of higher spin gravity: BPST
  instanton},'' \href{http://dx.doi.org/10.1007/JHEP07(2024)032}{{\em JHEP}
  {\bfseries 07} (2024) 032}, \href{http://arxiv.org/abs/2403.17148}{{\ttfamily
  arXiv:2403.17148 [hep-th]}}.

\bibitem{Sullivan77}
D.~Sullivan, ``Infinitesimal computations in topology,'' {\em Publ. Math. IHES}
  {\bfseries 47} (1977) 269--331.

\bibitem{vanNieuwenhuizen:1982zf}
P.~van Nieuwenhuizen, ``{Free graded differential superalgebras},'' in {\em
  {Group Theoretical Methods in Physics. Proceedings, 11th International
  Colloquium, Istanbul, Turkey, August 23-28, 1982}}, pp.~228--247.
\newblock
1982.
\newblock

\bibitem{DAuria:1980cmy}
R.~D'Auria, P.~Fre, and T.~Regge, ``{Graded Lie Algebra Cohomology and
  Supergravity},''
{\em Riv. Nuovo Cim.} {\bfseries 3N12} (1980) 1.

\bibitem{Vasiliev:1988sa}
M.~A. Vasiliev, ``Consistent equations for interacting massless fields of all
  spins in the first order in curvatures,''
{\em Annals Phys.} {\bfseries 190} (1989) 59--106.

\bibitem{Skvortsov:2022unu}
E.~Skvortsov and R.~Van~Dongen, ``{Minimal models of field theories: SDYM and
  SDGR},'' \href{http://dx.doi.org/10.1007/JHEP08(2022)083}{{\em JHEP}
  {\bfseries 08} (2022) 083}, \href{http://arxiv.org/abs/2204.09313}{{\ttfamily
  arXiv:2204.09313 [hep-th]}}.

\bibitem{Misuna:2024dlx}
N.~Misuna, ``{Unfolded formulation of 4d Yang-Mills theory},''
  \href{http://dx.doi.org/10.1016/j.physletb.2025.139882}{{\em Phys. Lett. B}
  {\bfseries 870} (2025) 139882},
  \href{http://arxiv.org/abs/2408.13212}{{\ttfamily arXiv:2408.13212
  [hep-th]}}.

\bibitem{Misuna:2024ccj}
N.~Misuna, ``{Scalar electrodynamics and Higgs mechanism in the unfolded
  dynamics approach},'' \href{http://dx.doi.org/10.1007/JHEP12(2024)090}{{\em
  JHEP} {\bfseries 12} (2024) 090},
  \href{http://arxiv.org/abs/2402.14164}{{\ttfamily arXiv:2402.14164
  [hep-th]}}.

\bibitem{Vasiliev:1986td}
M.~A. Vasiliev, ``Free massless fields of arbitrary spin in the de sitter space
  and initial data for a higher spin superalgebra,''
{\em Fortsch. Phys.} {\bfseries 35} (1987) 741--770.

\bibitem{Skvortsov:2022wzo}
E.~Skvortsov and Y.~Yin, ``{On (spinor)-helicity and bosonization in
  AdS$_{4}$/CFT$_{3}$},'' \href{http://dx.doi.org/10.1007/JHEP03(2023)204}{{\em
  JHEP} {\bfseries 03} (2023) 204},
  \href{http://arxiv.org/abs/2207.06976}{{\ttfamily arXiv:2207.06976
  [hep-th]}}.

\bibitem{Benincasa:2011pg}
P.~Benincasa and E.~Conde, ``{Exploring the S-Matrix of Massless Particles},''
  \href{http://dx.doi.org/10.1103/PhysRevD.86.025007}{{\em Phys. Rev. D}
  {\bfseries 86} (2012) 025007},
  \href{http://arxiv.org/abs/1108.3078}{{\ttfamily arXiv:1108.3078 [hep-th]}}.

\bibitem{BCF}
R.~Britto, F.~Cachazo, and B.~Feng, ``New recursion relations for tree
  amplitudes of gluons,''
  \href{http://dx.doi.org/10.1016/j.nuclphysb.2005.02.030}{{\em Nuclear Physics
  B} {\bfseries 715} no.~1–2, (May, 2005) 499–522}.
  \url{http://dx.doi.org/10.1016/j.nuclphysb.2005.02.030}.

\bibitem{BCFW}
R.~Britto, F.~Cachazo, B.~Feng, and E.~Witten, ``Direct proof of the tree-level
  scattering amplitude recursion relation in yang-mills theory,''
  \href{http://dx.doi.org/10.1103/physrevlett.94.181602}{{\em Physical Review
  Letters} {\bfseries 94} no.~18, (May, 2005) }.
  \url{http://dx.doi.org/10.1103/PhysRevLett.94.181602}.

\bibitem{Berends:1987me}
F.~A. Berends and W.~T. Giele, ``{Recursive Calculations for Processes with n
  Gluons},'' \href{http://dx.doi.org/10.1016/0550-3213(88)90442-7}{{\em Nucl.
  Phys. B} {\bfseries 306} (1988) 759--808}.

\bibitem{Krasnov:2016emc}
K.~Krasnov, ``{Self-Dual Gravity},''
  \href{http://dx.doi.org/10.1088/1361-6382/aa65e5}{{\em Class. Quant. Grav.}
  {\bfseries 34} no.~9, (2017) 095001},
  \href{http://arxiv.org/abs/1610.01457}{{\ttfamily arXiv:1610.01457
  [hep-th]}}.

\bibitem{chattopadhyay2023aspectsSDYM}
P.~Chattopadhyay, ``Aspects of self-dual yang-mills and self-dual gravity,''
  2023.
\newblock \url{https://arxiv.org/abs/2205.03675}.

\bibitem{Monteiro_2011}
R.~Monteiro and D.~O’Connell, ``The kinematic algebra from the self-dual
  sector,'' \href{http://dx.doi.org/10.1007/jhep07(2011)007}{{\em Journal of
  High Energy Physics} {\bfseries 2011} no.~7, (July, 2011) }.
  \url{http://dx.doi.org/10.1007/JHEP07(2011)007}.

\bibitem{Monteiro:2022xwq}
R.~Monteiro, ``{From Moyal deformations to chiral higher-spin theories and to
  celestial algebras},'' \href{http://dx.doi.org/10.1007/JHEP03(2023)062}{{\em
  JHEP} {\bfseries 03} (2023) 062},
  \href{http://arxiv.org/abs/2212.11266}{{\ttfamily arXiv:2212.11266
  [hep-th]}}.

\bibitem{Serrani:2025owx}
M.~Serrani, ``{On classification of (self-dual) higher-spin gravities in flat
  space},'' \href{http://dx.doi.org/10.1007/JHEP08(2025)032}{{\em JHEP}
  {\bfseries 08} (2025) 032}, \href{http://arxiv.org/abs/2505.12839}{{\ttfamily
  arXiv:2505.12839 [hep-th]}}.

\bibitem{Serrani:2025oaw}
M.~Serrani, ``{Associativity of celestial OPE, higher spins and
  self-duality},'' \href{http://arxiv.org/abs/2508.16804}{{\ttfamily
  arXiv:2508.16804 [hep-th]}}.

\bibitem{Skvortsov:2018uru}
E.~Skvortsov, ``{Light-Front Bootstrap for Chern-Simons Matter Theories},''
  \href{http://dx.doi.org/10.1007/JHEP06(2019)058}{{\em JHEP} {\bfseries 06}
  (2019) 058}, \href{http://arxiv.org/abs/1811.12333}{{\ttfamily
  arXiv:1811.12333 [hep-th]}}.

\bibitem{Jain:2024bza}
S.~Jain, D.~K. S, and E.~Skvortsov, ``{Hidden sectors of Chern-Simons matter
  theories and exact holography},''
  \href{http://dx.doi.org/10.1103/PhysRevD.111.106017}{{\em Phys. Rev. D}
  {\bfseries 111} no.~10, (2025) 106017},
  \href{http://arxiv.org/abs/2405.00773}{{\ttfamily arXiv:2405.00773
  [hep-th]}}.

\bibitem{Aharony:2024nqs}
O.~Aharony, R.~R. Kalloor, and T.~Kukolj, ``{A chiral limit for
  Chern-Simons-matter theories},''
  \href{http://dx.doi.org/10.1007/JHEP10(2024)051}{{\em JHEP} {\bfseries 10}
  (2024) 051}, \href{http://arxiv.org/abs/2405.01647}{{\ttfamily
  arXiv:2405.01647 [hep-th]}}.

\bibitem{Skvortsov:2026gtq}
E.~Skvortsov and R.~Van~Dongen, ``{Dirichlet, Neumann, Mixed and self-dual
  holography: (self-dual) Yang-Mills theory},''
  \href{http://arxiv.org/abs/2602.21658}{{\ttfamily arXiv:2602.21658
  [hep-th]}}.

\bibitem{Maldacena:2011nz}
J.~M. Maldacena and G.~L. Pimentel, ``{On graviton non-Gaussianities during
  inflation},'' \href{http://dx.doi.org/10.1007/JHEP09(2011)045}{{\em JHEP}
  {\bfseries 09} (2011) 045}, \href{http://arxiv.org/abs/1104.2846}{{\ttfamily
  arXiv:1104.2846 [hep-th]}}.

\bibitem{Lipstein:2023pih}
A.~Lipstein and S.~Nagy, ``{Self-Dual Gravity and Color-Kinematics Duality in
  AdS4},'' \href{http://dx.doi.org/10.1103/PhysRevLett.131.081501}{{\em Phys.
  Rev. Lett.} {\bfseries 131} no.~8, (2023) 081501},
  \href{http://arxiv.org/abs/2304.07141}{{\ttfamily arXiv:2304.07141
  [hep-th]}}.

\bibitem{Chowdhury:2024dcy}
C.~Chowdhury, G.~Doran, A.~Lipstein, R.~Monteiro, S.~Nagy, and K.~Singh,
  ``{Light-cone actions and correlators of self-dual theories in AdS$_{4}$},''
  \href{http://dx.doi.org/10.1007/JHEP01(2025)172}{{\em JHEP} {\bfseries 01}
  (2025) 172}, \href{http://arxiv.org/abs/2411.04172}{{\ttfamily
  arXiv:2411.04172 [hep-th]}}.

\bibitem{Tran:2021ukl}
T.~Tran, ``{Twistor constructions for higher-spin extensions of (self-dual)
  Yang-Mills},'' \href{http://dx.doi.org/10.1007/JHEP11(2021)117}{{\em JHEP}
  {\bfseries 11} (2021) 117}, \href{http://arxiv.org/abs/2107.04500}{{\ttfamily
  arXiv:2107.04500 [hep-th]}}.

\bibitem{Herfray:2022prf}
Y.~Herfray, K.~Krasnov, and E.~Skvortsov, ``{Higher-spin self-dual Yang-Mills
  and gravity from the twistor space},''
  \href{http://dx.doi.org/10.1007/JHEP01(2023)158}{{\em JHEP} {\bfseries 01}
  (2023) 158}, \href{http://arxiv.org/abs/2210.06209}{{\ttfamily
  arXiv:2210.06209 [hep-th]}}.

\bibitem{Tran:2022tft}
T.~Tran, ``{Toward a twistor action for chiral higher-spin gravity},''
  \href{http://dx.doi.org/10.1103/PhysRevD.107.046015}{{\em Phys. Rev. D}
  {\bfseries 107} no.~4, (2023) 046015},
  \href{http://arxiv.org/abs/2209.00925}{{\ttfamily arXiv:2209.00925
  [hep-th]}}.

\bibitem{Tran:2025uad}
T.~Tran, ``{Anomaly-free twistorial higher-spin theories},''
  \href{http://dx.doi.org/10.1016/j.geomphys.2025.105740}{{\em J. Geom. Phys.}
  {\bfseries 221} (2026) 105740},
  \href{http://arxiv.org/abs/2505.13785}{{\ttfamily arXiv:2505.13785
  [hep-th]}}.

\bibitem{Mason:2025pbz}
L.~Mason and A.~Sharma, ``{Chiral higher-spin theories from twistor space},''
  \href{http://arxiv.org/abs/2505.09419}{{\ttfamily arXiv:2505.09419
  [hep-th]}}.

\bibitem{Delfino_2015}
G.~Delfino, K.~Krasnov, and C.~Scarinci, ``Pure connection formalism for
  gravity: Feynman rules and the graviton-graviton scattering,''
  \href{http://dx.doi.org/10.1007/jhep03(2015)119}{{\em Journal of High Energy
  Physics} {\bfseries 2015} no.~3, (Mar., 2015) }.
  \url{http://dx.doi.org/10.1007/JHEP03(2015)119}.

\end{thebibliography}
\end{document}